\documentclass[12pt,letterpaper]{article}
\pdfoutput=1

\usepackage{graphicx,array}
\usepackage[export]{adjustbox}
\usepackage[dvipsnames]{xcolor}
\definecolor{darkblue}{rgb}{0,0.1,0.5}
\definecolor{darkgreen}{rgb}{0,0.5,0.2}
\definecolor{darkred}{RGB}{153,26,0}
\usepackage[normalem]{ulem}
\definecolor{seablue}{rgb}{0,0.2,0.6}
\usepackage{latexsym}
\usepackage{amssymb, amsmath}
\usepackage{bm}        
\usepackage[numbers,sort&compress]{natbib}
\usepackage{bm,relsize}
\usepackage{slashed}
\definecolor{viola}{RGB}{134,41,198}
\usepackage{mathrsfs}
\usepackage{hyperref} 
\hypersetup{
    colorlinks=true,       
    linkcolor=darkblue,          
    citecolor=BrickRed,        
    filecolor=magenta,      
    urlcolor=MidnightBlue           
}
\usepackage[all]{hypcap} 
\usepackage{subcaption}
\usepackage{multirow}
\usepackage{authblk}

\usepackage{tikz}
\usepackage{lipsum}

\newlength{\RoundedBoxWidth}
\newsavebox{\GrayRoundedBox}
\newenvironment{GrayBox}[1][\dimexpr\textwidth-4.5ex]%
   {\setlength{\RoundedBoxWidth}{\dimexpr#1}
    \begin{lrbox}{\GrayRoundedBox}
       \begin{minipage}{\RoundedBoxWidth}}%
   {   \end{minipage}
    \end{lrbox}
    \begin{center}
    \begin{tikzpicture}%
       \draw node[draw=black,fill=black!10,rounded corners,%
             inner sep=2ex,text width=\RoundedBoxWidth]%
             {\usebox{\GrayRoundedBox}};
    \end{tikzpicture}
    \end{center}}

\usepackage{tkz-euclide}
\usetikzlibrary{shapes.misc}
\usetikzlibrary{decorations.pathmorphing, decorations.pathreplacing, decorations.shapes}	
\tikzset{
    v/.style={decorate, decoration={snake, segment length=3mm, amplitude=0.75mm}, draw},
    f/.style={draw=black, postaction={decorate},
        decoration={markings,mark=at position .6 with {\arrow[very thick]{latex}}}},
    fb/.style={draw=black, postaction={decorate},
        decoration={markings,mark=at position .4 with {\arrowreversed[very thick]{latex}}}},
    fnar/.style={draw=black},
    g/.style={decorate, draw=black,
        decoration={coil,amplitude=3pt, segment length=3.5pt}},
    s/.style={dashed,draw=black, postaction={decorate},
        decoration={markings,mark=at position .55 with {\arrow[very thick]{latex}}}},
    sb/.style={dashed,draw=black, postaction={decorate},
        decoration={markings,mark=at position .55 with {\arrowreversed[draw=black,very thick]{latex}}}},
    snar/.style={dashed,draw=black,line width =1.25pt},
    cross/.style={cross out, draw=black, minimum size=2*(#1-\pgflinewidth), inner sep=0pt, outer sep=0pt},
cross/.default={3pt},
}
\usetikzlibrary{decorations.pathmorphing}

\tikzset{snake it/.style={decorate, decoration=snake}}
\usepackage{amsmath}

\usepackage[font=small,labelfont=bf,labelsep=period]{caption}

\usepackage{jheppub} 

\newcommand{\dd}{\text{d}}

\newcommand{\Tr}{\mathrm{Tr}}

\newcommand{\be}{\begin{equation}}
\newcommand{\ee}{\end{equation}}

\newcommand{\braket}[1]{\left\langle #1 \right\rangle}
\newcommand{\sbraket}[1]{\left[ #1 \right]}
\newcommand{\asbraket}[3]{\left\langle #1 \vert #2 \vert #3 \right]}
\newcommand{\sabraket}[3]{\left[ #1 | #2 | #3 \right\rangle}
\newcommand{\bra}[1]{ \left\langle #1 \right|}
\newcommand{\sbra}[1]{ \left[ #1 \right|}
\newcommand{\ket}[1]{ \left| #1 \right\rangle}
\newcommand{\sket}[1]{ \left| #1 \right]}
\newcommand{\order}[1]{ \mathcal{O}\left( #1 \right)}

\newcommand{\La}{\mathscr{L}}
\newcommand{\Amp}{\mathcal{A}}
\newcommand{\del}{\partial}
\newcommand{\dalpha}{\dot \alpha}

\newcommand{\al}[1]{\begin{align}\begin{aligned} #1 \end{aligned}\end{align}}

\title{\bf ALPs, the on-shell way}

\author[a,b]{Enrico Bertuzzo,}
\author[c,d,e]{Christophe Grojean,}
\author[a,c]{Gabriel M. Salla}

\affiliation[a]{\it Instituto de F\'isica, Universidade de S\~ao Paulo, C.P. 66.318, 05315-970 S\~ao Paulo, Brazil}
\affiliation[b]{\it Dipartimento di Scienze Fisiche, Informatiche e Matematiche, Università di Modena e Reggio
Emilia, via G. Campi 213/A, 41125 Modena, Italy}
\affiliation[c]{\it Deutsches Elektronen-Synchrotron DESY, Notkestr. 85, 22607 Hamburg, Germany}
\affiliation[d]{\it Institut für Physik, Humboldt-Universität zu Berlin, 12489 Berlin, Germany}
\affiliation[e]{\it Theoretical Theoretical Physics Department, CERN, 1211 Geneva 23, Switzerland}

\abstract{We study how the coupling between axion-like particles (ALPs) and matter can be obtained at the level of on-shell scattering amplitudes. 
We identify three conditions that allow us to compute amplitudes that correspond to shift-symmetric Lagrangians, at the level of operators with dimension 5 or higher, and we discuss how they relate and extend the Adler's zero condition. These conditions are necessary to reduce the number of coefficients consistent with the little-group scaling to the one expected from the Lagrangian approach.  
We also show how our formalism easily explains that the dimension-5 interaction involving one ALP and two massless spin-1 bosons receive corrections from higher order operators only when the ALP has a non-vanishing mass. 
As a direct application of our results, we perform a phenomenological study of the inelastic scattering $\ell^+\ell^- \to  \phi h$ (with $\ell^\pm$ two charged leptons, $\phi$ the ALP and $h$ the Higgs boson) for which, as a result of the structure of the 3-point and 4-point amplitudes, dimension-7 operators can dominate over the dimension-5 ones well before the energy reaches the cutoff of the theory.}

\begin{document}
\begin{flushright}
CERN-TH-2023-202\\
DESY-23-090\\
HU-EP-23/21
\end{flushright}
\maketitle
\flushbottom

\section{Introduction}

In recent years, the study of modern on-shell methods\,\cite{Dixon:2013uaa,Elvang:2013cua,Arkani-Hamed:2017jhn,Travaglini:2022uwo, Henn:2014yza}, together with their application to phenomenological issues, has been gaining much attention and giving fruitful results. Without a doubt, the most innovative feature of these methods consists of writing down scattering amplitudes by relying on nothing but the covariance of the $S$-matrix under little-group transformations of the Lorentz group\,\cite{Weinberg:1995mt}, thus putting aside the need for fields and Lagrangians. The consequences and applications of the on-shell approach are far-reaching. On the phenomenological side, much progress has been made, for instance, in the computation of loop-integrals and anomalous dimensions\,\cite{Bern:1994cg,Abreu:2022mfk,Blumlein:2022zkr,Baratella:2020lzz,EliasMiro:2020tdv,Jiang:2020mhe,Baratella:2020dvw}, in the understanding of the Standard Model (SM) and of Effective Field Theories (EFTs)\,\cite{Cheung:2014dqa,Cheung:2015ota,Christensen:2022nja,Christensen:2018zcq,Christensen:2019mch,Aoude:2019tzn,Durieux:2019siw,Arkani-Hamed:2020blm,Balkin:2021dko,Bonnefoy:2021qgu,Durieux:2019eor,Machado:2022ozb,AccettulliHuber:2021uoa,DeAngelis:2022qco,Liu:2023jbq,Shadmi:2018xan,Dong:2021yak,Ma:2019gtx,Durieux:2020gip,DeAngelis:2023bmd}, in the study of the physics of higher-spin dark-matter\,\cite{Falkowski:2020fsu,Salla:2022dxc} and also in the formulation of neutrino oscillations\,\cite{Alves:2021rjc}. One less pursued question is that of establishing a precise connection between the physical properties of infra-red (IR) on-shell amplitudes to the physics of the ultra-violet (UV)\,\cite{Cheung:2014dqa,Cheung:2018oki,Dai:2020cpk, Liu:2023jbq,Balkin:2021dko,Arkani-Hamed:2008owk,Cheung:2016drk}. In the Lagrangian approach, the different assumptions about the UV dynamics are translated in the IR to specific EFTs and power counting (\textit{e.g.} SILH\,\cite{Giudice:2007fh}, HEFT\,\cite{Feruglio:1992wf}), giving us much more control over the properties of the low-energy amplitudes. The same exercise still needs to be carried out in a systematic way in the on-shell approach, in which the UV properties are reconstructed from the IR amplitudes.

Along these lines, an interesting problem that can be studied using on-shell methods is the one concerning the physics of axion-like particles (ALPs). From the usual quantum field theoretical perspective, it is well known that ALP interactions must be invariant under a shift-symmetry if the underlying global symmetry is exact. This has extensive physical consequences, among which the existence of \textit{soft-theorems} and the appearance of the so-called \textit{Adler's zero}, which state that amplitudes involving ALPs (or, equivalently, Nambu--Goldstone bosons) and amplitudes involving ALPs and other particles are either regular or vanish in the limit in which the ALP momentum becomes soft\,\cite{Adler:1964um,Weinberg:1996kr,Low:2017mlh,Low:2015ogb}. Previous studies in the literature have shown how, using on-shell techniques, amplitudes involving only exactly massless ALPs make manifest the previously mentioned soft-theorems\,\cite{Kampf:2019mcd}, as well as Adler's zero conditions\,\cite{Low:2018acv,Cheung:2016drk,Arkani-Hamed:2008owk,Low:2019ynd,Du:2015esa}. In addition, it is possible to read off from IR properties of these ALPs amplitudes the structure of the coset group associated to the spontaneous symmetry breaking in the UV\,\cite{Dai:2020cpk}. 

Two questions that have remained unanswered in the literature are: \textit{what happens when we consider couplings between ALPs and matter? What are the physical properties that these amplitudes must satisfy in order to recover shift-symmetry?} Stated in another way: the Adler's zero condition is a necessary requirement that an amplitude involving ALPs must satisfy, but it may not be enough to completely characterize the interactions between ALPs and matter. What we seek to find in this work are possible additional conditions that completely pinpoint the ALP-matter interactions. 
Given that on-shell methods allow us to write the amplitudes without assuming any Lagrangian or symmetry, they are the ideal framework to approach this question. 

In this paper, we continue to further explore this direction and investigate the coupling between ALPs and other matter particles, including the SM ones. We will first reproduce the well-known results of ALP 3-point couplings to massive fermions and vectors, which in the Lagrangian approach emerge from operators of the form $(\partial_\mu \phi) \bar{\psi} \gamma^\mu \gamma^5 \psi$ and $\phi V \tilde{V}$, where $\phi$ is the ALP, $\psi$ any massive fermion, $V$ a field strength tensor and $\tilde{V}$ denotes the dual field strength (to be defined below). 
We will then uplift these amplitudes to 3- and 4-point amplitudes involving SM particles in the unbroken electroweak phase. Understanding the properties of the simplest scattering amplitudes of ALPs and matter particles will allow us to generalize this procedure and construct higher-point functions in a systematic way.

The paper is organized as follows. In Section~\ref{sec:toy}, we formulate our approach in terms of on-shell methods and apply it to generic 3-point amplitudes involving one ALP. We discuss how we could generalize our procedure to higher-point amplitudes involving more ALPs and the difficulties involved.
In Section~\ref{sec:SM}, we match the massive amplitudes in the IR to the massless ones in the UV, while specializing to the SM particle content, discussing also how to handle electroweak symmetry breaking effects. We also comment on the physical interpretation of the shift-symmetry breaking invariants introduced in Ref.\,\cite{Bonnefoy:2022rik}. In Section~\ref{sec:higher}, we build higher-point contact amplitudes up to dimension 8 in the ALP scale, while some phenomenological applications are discussed in Section~\ref{sec:pheno}. In particular, we study the production of an ALP in association with a Higgs in a lepton collider, $\ell^-\ell^+\to \phi h$, and show that higher-order contact operators can give the leading contribution to the cross-section at high-energies. Finally, we conclude in Section~\ref{sec:conclusion}. We also add a number of appendices with more technical material: in Appendix~\ref{app:conventions}, we present our conventions for spinors; in Appendix~\ref{app:Yukawa-mass}, we propose an alternative on-shell derivation of the connection between Yukawa couplings and fermions masses; finally, in Appendix~\ref{app:running}, we present detailed computations of the running of the 4-particle amplitude involving one ALP, a fermion-antifermion pair and one Higgs doublet. This will be useful to show the consistency between the 3-point amplitude involving one ALP and two massive fermions and the 3-point amplitude involving one ALP and two massless gauge bosons.

\section{ALP couplings to matter}\label{sec:toy}

\subsection{General remarks}

On-shell techniques have been previously used in the literature to study amplitudes involving massless scalar particles. More precisely, under the assumption that these amplitudes vanish as any of the momenta goes soft, \textit{i.e.} $p\to 0$, it is possible to derive a number of features of such (pseudo-)scalars and in some cases even completely determine the underlying theory\,\cite{Cheung:2014dqa,Cheung:2018oki,Kampf:2019mcd,Low:2018acv,Cheung:2016drk,Arkani-Hamed:2008owk,Dai:2020cpk}. However, these analyses are restricted to amplitudes with nothing but scalars, and therefore do not apply when they interact with other matter fields, for instance SM particles, making necessary the addition of extra conditions to characterise ALPs amplitudes. One of the goals of the present paper is then to extend this discussion and to characterise the interactions of ALPs with other particles from an on-shell perspective.

Our starting point are amplitudes involving ALPs, hereafter denoted by $\phi$, in the limit in which the ALP momentum, $p_\phi$, goes to zero. We will thus be focusing on 
\be\label{eq:lim_amp}
\lim_{p_\phi\to 0}\Amp\!\left[\phi,~ \mathcal{O}\right],
\ee
where $\Amp$ denotes the amplitude and $\mathcal{O}$ is a set of other arbitrary particles.\footnote{A more precise definition of the soft limit in terms of Lorentz invariant quantities is $|p_\phi\cdot p_i|\ll |p_i\cdot p_j|$, for any two momenta $p_i\neq p_j$ of the set $\cal O$.} We will start by considering amplitudes with the minimal number of particles coupled to one ALP (taken to be massless), \textit{i.e.} 3-point amplitudes $\Amp_3$ in which one ALP interacts with two other particles. Our purpose is to use the soft limit to deduce which conditions such 3-point amplitudes must satisfy. We will discuss how to do this in a concrete way shortly. To avoid possible complications due to SM symmetry group, in this section ``matter'' means any massive particle of spin 0, 1/2 or 1 that can couple to the ALP.
We will see in Section\,\ref{sec:SM} that our results can be directly applied to amplitudes of massive SM particles (that would correspond to the broken electroweak phase), and we will also discuss how to extend them to the massless SM particles (above the electroweak scale).

In on-shell language, 3-point amplitudes are special objects, since they are completely fixed by little-group covariance and have constant coefficients. While massless 3-point amplitudes only make sense using complex momenta (the 3-body kinematics forces the amplitude to vanish for real momenta)\,\cite{Elvang:2013cua,Henn:2014yza}, no such obstruction arise for massive amplitudes and the momenta can be taken real. 
Also, 3-point amplitudes can be used as building blocks to form the so-called ``constructible'' higher-point amplitudes\,\cite{Britto:2004ap,Britto:2005fq,AccettulliHuber:2021uoa}. Contact interactions, that cannot be constructed in this way, will be discussed in Section~\ref{sec:higher}. We will now focus on the coupling between one ALP and two massive particles of the same species (we will relax this condition in Section \ref{sec:toy_many}). The amplitudes we are interested in are written as
\be\label{eq:3_pointamp_toy}
\Amp_3\!\left[\phi \,\mathcal{P}^{I_1,\cdots, I_{2s}}_1  \mathcal{\bar P}^{J_1,\cdots, J_{2s}}_2 \right],
\ee
where $\mathcal{P}$ is a particle of spin $s$, $\bar{\mathcal{P}}$ its antiparticle, $\{I_1,\cdots I_{2s}\},~\{J_1,\cdots, J_{2s}\}$ are symmetrized massive little-group indices and the subscripts denote the labels of the momenta (for more details on the notation and conventions, we refer the reader to Appendix~\ref{app:conventions}).

\begin{figure}[t]
\centering
\begin{equation*}
\lim_{p_\phi\to 0}~~~
\adjustbox{valign=m}{\begin{tikzpicture}[line width=0.75]
\draw[fill=white] (0,0) circle (20pt) node[midway] {$\Amp_{n+1}$};
\draw (1-0.49,0.49) -- (2-0.49,1) node[right] {$\mathcal{P}_{i,p}$};
\draw[dashed] (1-0.49,-0.49) -- (2-0.49,-1) node[right] {$\phi$};
\draw (-1+0.49,0.49) -- (-2+0.49,1);
\draw (-1+0.49,-0.49) -- (-2+0.49,-1);
\filldraw[black] (-1.3,0) circle (0.75pt);
\filldraw[black] (-1.2,0.3) circle (0.75pt);
\filldraw[black] (-1.2,-0.3) circle (0.75pt);
\filldraw[black] (1.3,0) circle (0.75pt);
\filldraw[black] (1.2,0.3) circle (0.75pt);
\filldraw[black] (1.2,-0.3) circle (0.75pt);
\end{tikzpicture}}
=\lim_{p_\phi\to 0}\sum_{i=1}^n~~
\adjustbox{valign=m}{\begin{tikzpicture}[line width=0.75]
\draw[fill=white] (0,0) circle (16pt) node[midway] {$\Amp_{n}$};
\draw (-1+0.55,0.35) -- (-1.3,0.7);
\draw (-1+0.55,-0.35) -- (-1.3,-0.7);
\filldraw[black] (-1.2,0) circle (0.75pt);
\filldraw[black] (-1.1,0.25) circle (0.75pt);
\filldraw[black] (-1.1,-0.25) circle (0.75pt);
\draw (0.55,0) -- (2,0) node[above,midway] {$\mathcal{P}_{i,p+p_\phi}$};
\draw[fill=white] (2.55,0) circle (16pt) node(2,55,0) {$\Amp_{3}$};
\draw (1-0.55+2.55,0.35) -- (1.3+2.55,0.7) node[right] {$\mathcal{P}_{i,p}$};
\draw[dashed] (1-0.55+2.55,-0.35) -- (1.3+2.55,-0.7) node[right] {$\phi$};
\end{tikzpicture}}
\end{equation*}
\caption{Diagramatic representation of Eq.\,\eqref{eq:soft_factorizarion}. The index $i$ labels the particle species. The momenta of external particles are taken to be incoming.}
\label{fig:soft_factorization}
\end{figure}
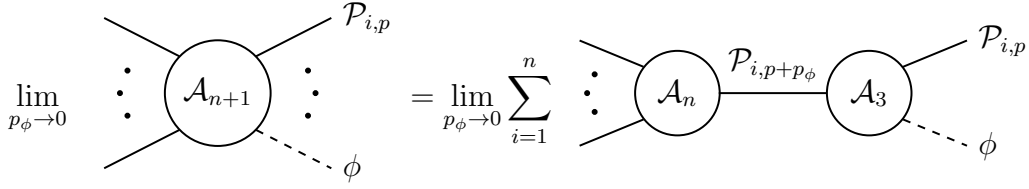

Going back to Eq.\,\eqref{eq:lim_amp}, it is important to notice that the soft limit effectively ``freezes'' the particle whose momentum is becoming soft, leaving us with an amplitude with one less dynamical particle. This means that, when we try to apply \eqref{eq:lim_amp} to 3-point amplitudes as in Eq.\,\eqref{eq:3_pointamp_toy}, the soft limit would leave us with a non-physical 2-point amplitude. To circumvent this obstacle, we will consider the 3-point amplitude in Eq.\,\eqref{eq:3_pointamp_toy} as part of a generic $(n+1)$-point amplitude $\Amp_{n+1}\!\left[\phi,\mathcal{P},\mathcal{O}\right]$, constructed ``joining'' together $\Amp_3\!\left[ \phi \mathcal{P}\bar{\mathcal{P}}\right]$ in Eq.\,\eqref{eq:3_pointamp_toy} with a $n$-point amplitude $\Amp_n\!\left[\mathcal{P},\mathcal{O}\right]$ (see Fig.\,\ref{fig:soft_factorization}). The additional particles denoted collectively as $\mathcal{O}$ are generic (apart from the fact that there must not be another ALP) and their nature will not play any role in our reasoning.\,\footnote{We will come back to the case with more ALPs in the end of this Section and in Section \ref{sec:higher}.} In the soft limit, the $(n+1)$-point amplitude decomposes as
\begin{multline}\label{eq:soft_factorizarion}
    \lim_{p_\phi\to 0}  \Amp_{n+1}\!\left[\phi\,\mathcal{P}^{I_1,\cdots, I_{2s}}_{i,p},\cdots \right] = \\
    \lim_{p_\phi\to 0}\sum_{i=1}^n \Amp_n\!\left[\mathcal{P}^{K_1,\cdots ,K_{2s}}_{i,p+p_\phi},\cdots\right] \frac{\epsilon_{K_1 J_1}\cdots\epsilon_{ K_{2s}J_{2s}}}{(p+p_\phi)^2-m_{\mathcal{P}_i}^2}\Amp_3\!\left[\phi \mathcal{P}^{I_1,\cdots, I_{2s}}_{i,p}  \mathcal{\bar P}^{J_1,\cdots, J_{2s}}_{i,-p-p_\phi}\right],
\end{multline}
where we sum over all particles $\mathcal{P}$,
which are labeled by $i$ in $\Amp_{n+1}$ and $\Amp_n$, and $\epsilon_{IJ}$ is the Levi--Civita tensor that takes into account the sum over spin configurations of the propagating particle (see Fig.~\ref{fig:soft_factorization}). We can understand Eq.\,\eqref{eq:soft_factorizarion} as follows. Since the ALP couples to two particles of the same species, in the $p_\phi \to 0$ limit the momentum of the particle in the propagator is very close to $p^2 = m_{\mathcal{P}_i}^2$, \textit{i.e.} the particle is very close to its mass shell. According to {\it polology}\,\cite{Weinberg:1995mt},\,\footnote{As the name suggests, \textit{polology} refers to the pole structure of amplitudes. More precisely, suppose we consider a $n$-point amplitude $\mathcal{A}_n$ in which the external momenta $p_{1, \dots, r}$ and $p_{r+1, \dots, n}$ (for some $r<n$) are such that $s_{1,\dots,r} \equiv (p_1 + \dots + p_r)^2 = m^2$, where $m$ is the mass of some 1-particle state $\psi$ that can be exchanged in an internal propagator. If this is the case, then \textit{polology} guarantees that the amplitude has a pole at $s_{1,\dots,r} = m^2$ and that the residue factorizes into the product of the two subamplitudes connected by the propagator of the particle $\psi$:
$$
\lim_{s_{1,\dots,r} \to m^2}{\rm Res} (\mathcal{A}_n) = \mathcal{A}_{r+1}\left(p_1\dots p_r p_\psi\right) \mathcal{A}_{n-r+1}\left(p_{r+1} \dots p_n (-p_\psi)\right),
$$
where $-p_\psi = p_1 + \dots + p_r =- p_{r+1} - \dots - p_n$ is given by momentum conservation (we are taking all momenta incoming). This result is non-perturbative and relies only on the locality and unitarity of the $S$-matrix. Furthermore, it is the basis of the Britto--Cachazo--Feng--Witten (BCFW) recursion relations\,\cite{Britto:2004ap,Britto:2005fq}.} the total amplitude will then factorize into the product of the two sub-amplitudes multiplied by the intermediate propagator and hence we obtain Eq.\,\eqref{eq:soft_factorizarion}. We can rewrite it more compactly as
\be\label{eq:soft_limit_toy}
\lim_{p_\phi\to 0}  \Amp_{n+1}\!\left[\phi\,\mathcal{P}^{I_1,\cdots, I_{2s}}_{i,p},\cdots\right] =\lim_{p_\phi\to 0}  \sum_{i=1}^n \Amp_n\!\left[\mathcal{P}^{K_1,\cdots, K_{2s}}_{i,p}, \cdots\right] \times (\mathcal{S}_{\mathcal{P}_i})_{K_1,\cdots, K_{2s}}^{I_1,\cdots, I_{2s}},
\ee
with the soft factor given by
\be\label{eq:soft_factor_toy}
(\mathcal{S}_{\mathcal{P}_i})_{K_1,\cdots,  K_{2s}}^{I_1,\cdots, I_{2s}} 
 = \frac{\epsilon_{K_1 J_1}\cdots\epsilon_{ K_{2s}J_{2s}}}{2p\cdot p_\phi}\Amp_3\!\left[\phi \mathcal{P}^{I_1,\cdots, I_{2s}}_{i,p}  \mathcal{\bar P}^{J_1,\cdots, J_{2s}}_{i, -p-p_\phi}\right].
\ee
In the last step we have simplified the propagator using the fact that $\phi$ is massless, which gives $(p+p_\phi)^2-m_{\mathcal{P}_i}^2 = 2p\cdot p_\phi$. We stress once more that all the particles appearing in the soft factor are on-shell and that we consider real momenta. We observe that, under our hypothesis, the amplitude $\mathcal{A}_{n+1}$ has the same kinematical configuration of $\mathcal{A}_n$ without the ALP, \textit{i.e.} the ingoing particle $\mathcal{P}_i$ is on-shell in both cases.
We are therefore led to the conclusion that the amplitude describing the interactions of an ALP with matter particles should obey the following soft factorization condition:
\begin{GrayBox}
\textbf{ALP soft factorization condition}
\vspace{0.5em}

\textit{
For a massless ALP, in the limit $p_\phi\to0$, as the amplitude $\Amp_{n+1}\!\left[\phi,\mathcal{P},\mathcal{O}\right]$ factorizes as}
\be\label{eq:soft_factor}
\Amp_{n+1}\!\left[\phi,\mathcal{P},\mathcal{O}\right]\xrightarrow{p_\phi\to 0} \sum_{\mathcal{P}} \Amp_n\!\left[\mathcal{P},\mathcal{O}\right]\times \mathcal{S_P},
\ee
\textit{
we demand that no poles appear in $\mathcal{S_P}$ in \textbf{all} phase space for real momenta. This condition holds when the 3-point amplitude $\mathcal{A}_3$ that appears in $\mathcal{S_P}$ represents the interaction of the ALP with two particles of the same species, such that $\mathcal{A}_{n+1}$ has the same kinematical structure as $\mathcal{A}_n$ without the ALP.
}
\end{GrayBox}
The motivation for the condition in Eq.\,\eqref{eq:soft_factor} comes from the fact that we want the regularity of the soft-limit to be a general property of the amplitude, and not simply a characteristic of a particular point in the phase-space. Stated in another way: given the factorization in Eq.\,\eqref{eq:soft_limit_toy}, the fact that particles $\mathcal{P}$ and $\mathcal{O}$ are the same in $\Amp_n$ and $\Amp_{n+1}$ guarantees that the condition of regularity will be valid for any kinematical configuration of $\mathcal{P}$, $\mathcal{O}$ and will thus be independent of particular choices of momenta.
Due to the singularities of the propagator in $\mathcal{S}_\mathcal{P}$, the ALP soft factorization condition is expected to give non-trivial requirements on $\Amp_3$.\footnote{We see that the factorization in Eq.\,\eqref{eq:soft_limit_toy} may in general affect the spin configuration of the remaining $n$-point amplitude when compared to the original $(n+1)$-point one. As we will see, however, the leading term in the soft expansion will always satisfy $(\mathcal{S}_\mathcal{P})_{K_1,\cdots, K_{2s}}^{I_1,\cdots, I_{2s}}\propto \delta^{I_1}_{K_1}\cdots\delta^{I_{2s}}_{K_{2s}}+\text{symm}$, {\it i.e.} the spin configuration will remain the same, with a change in spin configuration arising only in the subleading terms. Here, "symm" indicates the symmetric combinations of little-group indices.}

To make progress we need to specify the particle content of the amplitude $\Amp_3$.
We will now analyze in turn the cases in which ${\cal P}$ has spin 0, 1/2 and 1.

\subsection{Scalars}
In the simple case in which $\mathcal{P}=S$, with $S$ a spin-0 particle, there are no little group indices associated with $S$ and $\mathcal{S}_S$ is simply given by
\be\label{eq:S_s}
\mathcal{S}_S = \frac{1}{2p\cdot p_\phi}\Amp_3\!\left[\phi S_p \bar S_{-p-p_\phi}\right].
\ee
The 3-point amplitude among 3 scalars amounts just to a simple constant, therefore the only way to avoid $\mathcal{S}_S$ of diverging as $p_\phi\to 0$ is to set it to zero. This is nothing but a manifestation of the fact that in the usual quantum theoretical language the interactions of an ALP with two scalars, given by $(\del_\mu \phi) (S^\dagger i\overleftrightarrow{\del_\mu} S)$, gives a vanishing amplitude when all particles are taken on-shell.

\subsection{Fermions}\label{subsec:fermions}

Moving to the case of fermions $\mathcal{P}=\psi$, we now have a non-trivial little-group scaling and the 3-point amplitude can be written as
\be\label{eq:ALP_2fermions_toy}
\Amp_3\!\left[\phi\psi_1^I\bar \psi_2^J\right] = g_L \braket{\bm{12}} + g_R \sbraket{\bm{12}},
\ee
with two a priori complex dimensionless constants, $g_{L,R}$, that are related to the couplings to the left- and right-handed components of the fermions, respectively. We consider for concreteness the case of Dirac fermions, but the same discussion holds for Majorana fermions. In the previous amplitude we have used the {\it bold} notation introduced in Ref.\,\cite{Arkani-Hamed:2017jhn}, which amounts to simply bold the particle momenta instead of writing explicitly the little group indices. More specifically, in this case the bolded spinor products are matrices in the $\left\{I,J \right\}$ space (see Appendix~\ref{app:conventions}).
Also, we use the short-hand notation $\ket{\bm{n}}\equiv \ket{\bm{p_n}}$, and similar for other spinors, to label the momenta. Calling $p_1=p$, $p_2=-p-p_\phi$ and using Eqs.\,\eqref{eq:equal_braket} and \eqref{eq:analytic_cont}, the soft factor $\mathcal{S}_\psi$ becomes
\be\label{eq:S_psi}
(\mathcal{S}_\psi)^I_K 
= \frac{m_\psi}{2p\cdot p_\phi}(g_L+g_R)\delta^I_K + (\mathcal{S}_{p_\phi^0})^I_K,
\ee
where $\mathcal{S}_{p_\phi^0}$ denotes terms that are sub-leading
on $p_\phi$ and we will always assume masses to be real and positive. As we are going to see, both of these terms will play an important role in determining the form of $g_L$ and $g_R$.
First, to avoid the divergence of $\mathcal{S}_\psi$ 
in the soft limit
we need $g_L=-g_R$,\footnote{Since, under parity, angle and square brackets are exchanged, the condition $g_L = - g_R$ amounts to a parity-odd amplitude, \textit{i.e.} to a pseudoscalar ALP. We thank Yael Shadmi for pointing this out.} which implies that the coupling of the ALP to the fermions must necessarily be axial. This condition is nevertheless not enough to guarantee the regularity for all $p$, as a divergence may still appear when the (real) 3-momenta $\vec{p}$ and $\vec{p}_\phi$ are collinear with $\vec{p}\cdot\vec{p}_\phi>0$. In this case, the denominator appearing in Eq.\,\eqref{eq:S_psi}, that is contained in the definition of $\mathcal{S}_{p_\phi^0}$, becomes
\be\label{eq:high_energy_collinear}
\frac{1}{2p\cdot p_\phi} = \frac{1}{2|\vec{p}_\phi|\left(\sqrt{m_\psi^2+|\vec{p}\,|^2} - |\vec{p}\,|\right)}\xrightarrow{\frac{m_\psi}{|\vec{p}|}\ll 1}\frac{|\vec{p}\,|}{|\vec{p}_\phi| m_\psi^2},
\ee
\textit{i.e.} it diverges for very small masses. We anticipate here that the massless limit $m_\psi \to 0$ is precisely the limit that will allow us in Section\,\ref{sec:SM} to match the massive SM amplitudes in the IR into the corresponding massless amplitudes invariant under the SM symmetry group in the UV.
To avoid divergencies when $m_\psi/|\vec{p}| \ll 1$, we must then require the soft factor ${\cal S}_\psi$ to be regular also in the collinear configuration.
Since, with the condition $g_\psi=g_L = -g_R$, the first term in Eq.\,\eqref{eq:S_psi} vanishes, we turn to $\mathcal{S}_{p_\phi^0}$. In the collinear limit, this term is proportional to (see Eq.\,\eqref{eq:collinear_product_massive})
\be
\lim_{\rm collinear} (\mathcal{S}_{p_\phi^0})^I_K = -\frac{g_\psi\, \epsilon_{KJ}}{2 m_\psi} \left( \delta^I_2 \delta^J_1 + \delta^J_2 \delta^I_1 \right) ,
\ee
where the structure carrying the little group indices is diagonal and traceless because of the contraction with the $\epsilon_{KJ}$ tensor, $-\epsilon_{KJ}  \left(\delta^I_2 \delta^J_1 + \delta^J_2 \delta^I_1 \right) = (\sigma^3)^I_K$.
The only way to guarantee a non-singular massless limit is to have the couplings to be proportional to the mass, since then Eq.\,\eqref{eq:S_psi} becomes independent of $m_\psi$ in this limit. As a consequence, remembering that the couplings $g_{L,R}$ are dimensionless, we are forced to introduce a new scale $f$ in order to correct their dimensionality. From the arguments above, we then find that $g_R=-g_L\propto m_\psi/f$.

We can yet obtain more information about the phase of the couplings by imposing $CPT$ invariance and unitarity of the amplitude. The relation of the amplitude with their $CPT$ conjugate is given by Eq.\,\eqref{eq:CPT_amp}, and implies in the present case that $g_L^{\phantom{*}}=g_R^*$. This, together with the previous conditions, leads to purely imaginary coefficients. In short, we conclude that the coefficients in Eq.\,\eqref{eq:ALP_2fermions_toy} must be of the form
\be\label{eq:coupling_toy_fermion}
g_L=-g_R = C_{\phi\psi\psi}\frac{im_\psi}{f},\quad C_{\phi\psi\psi}\in \mathbb{R}.
\ee
The non-trivial generalisation to several fermion species will be presented in Eq.~(\ref{eq:coupling_toy_fermion_many}).
With Eq.\,\eqref{eq:coupling_toy_fermion}, the final expression for the amplitude then reads
\be\label{eq:ALP_2fermions_final}
\Amp_3\!\left[\phi\psi_1^I\bar\psi_2^J\right] = C_{\phi\psi\psi}\frac{im_\psi}{f}\big(\braket{\bm{12}}-\sbraket{\bm{12}}\big).
\ee
This result agrees exactly with what we have from the usual quantum theoretical approach starting from an interaction given by $(C_{\phi\psi\psi}/2f)(\del_\mu\phi)\bar \psi \gamma^5 \gamma^\mu \psi$. We do not have any coupling with the vector current, because it is not physical due to vector current conservation on-shell. It is worth stressing that, while little-group arguments were predicting the structure of the 3-point amplitude in term of two complex constants, the ALP soft factorization condition \eqref{eq:soft_factor} reduces this freedom to a single real parameter.

\subsection{Vectors}

We now move to the case $\mathcal{P}=V$, with $V$ a spin 1 particle of mass $m_V$. Without any loss of generality we take $\bar V = V$. The 3-point amplitude reads
\be\label{eq:ALP_2vectors_toy}
\Amp_3\!\left[\phi V_1^{I_1,I_2}V_2^{J_1,J_2}\right] = \frac{g_-}{f}\braket{\bm{12}}^2+\frac{g_+}{f}\sbraket{\bm{12}}^2 + \frac{g_0}{m_V}\braket{\bm{12}}\sbraket{\bm{12}},
\ee
where $g_\pm,g_0$ are three dimensionless complex constants and $f$ is again a new scale needed to correct the dimension of the coefficients.
We observe that, unlike $g_\pm$, the coupling $g_0$ is instead divided by a factor $m_V^{-1}$ to ensure a well-defined high-energy limit\,\cite{Liu:2023jbq,Durieux:2019eor}. Similar to Eq.\,\eqref{eq:ALP_2fermions_toy}, we use the bold notation to leave the little-group indices implicit. It is worth stressing again that we need to symmetrize over the little-group indices, so for instance\,\footnote{Here we choose the normalization of 1/2 for all values of $I_{1,2},J_{1,2}$ for simplicity, but other conventions can be useful in other contexts, for example in Refs.\,\cite{Christensen:2019mch,Durieux:2019eor}.}
\be
\braket{\bm{12}}^2 = \frac{1}{2}\left(\braket{1^{I_1}2^{J_1}}\braket{1^{I_2}2^{J_2}}+\braket{1^{I_2}2^{J_1}}\braket{1^{I_1}2^{J_2}}\right),
\ee
and analogous expressions for the other spinor structures. With this amplitude we compute the soft factor:
\al{\label{eq:S_V}
(\mathcal{S}_V)^{I_1,I_2}_{K_1,K_2}= \frac{m_V^2}{2p\cdot p_\phi}\left(\frac{g_-+g_+}{f} + \frac{g_0}{m_V}\right)\frac{1}{2}\left(\delta^{I_1}_{K_1}\delta^{I_2}_{K_2}+\delta^{I_2}_{K_1}\delta^{I_1}_{K_2}\right) + (\mathcal{S}_{p_\phi^0})^{I_1, I_2}_{K_1, K_2},
}
where again $\mathcal{S}_{p_\phi^0}$ are terms that are subleading in $p_\phi$. 
As in the case of the fermions, regularity when $|\vec{p}_\phi|\to 0$ imposes that the 
term in brackets of the first term vanishes. Regularity in the collinear limit (when $m_V \to 0$) requires $\mathcal{S}_{p_\phi^0}$ to be regular, and we can easily show that this requires $g_0=0$ (recall Eq.\,\eqref{eq:high_energy_collinear}). Together with $CPT$ and unitarity, the constraints on the couplings we obtain are
\be\label{eq:coupling_toy_vector}
g_-=-g_+ = iC_{\phi VV},\quad C_{\phi VV}\in \mathbb{R}.
\ee
Similarly to the case of fermions, the equation above implies that the couplings of the ALP to two vectors are purely imaginary and axial. With the results above, the amplitude in Eq.\,\eqref{eq:ALP_2vectors_toy} becomes
\be\label{eq:ALP_2vectors_toy_final}
\Amp_3\!\left[\phi V_1^{I_1,I_2}V_2^{J_1,J_2}\right] = \frac{iC_{\phi VV}}{f}\left(\braket{\bm{12}}^2-\sbraket{\bm{12}}^2\right).
\ee
It is possible to show that the amplitude with such couplings corresponds exactly to the operator $(C_{\phi VV}/2f)\phi V_{\mu\nu}\tilde V^{\mu\nu}$, where $\tilde V_{\mu\nu}\equiv \frac{1}{2} \epsilon_{\mu\nu\alpha\beta}V^{\alpha\beta}$ (see Appendix~\ref{app:conventions}). Once again, we see that the ALP soft factorization condition \eqref{eq:soft_factor} reduces the number of free parameters from three complex to one real coupling.

\subsection{Fermions - Many species case}\label{sec:toy_many}

So far we have analysed the couplings of the ALP assuming that only one species of particle $\mathcal{P}$ couples to $\phi$. For the discussion in Section~\ref{sec:SM}, it will also be necessary to consider the case in which the ALP couples to at least two non-degenerate fermion species. What we would like to do is to generalize what has been done in Section\,\ref{sec:toy}, constructing a $(n+1)$-amplitude $\Amp_{n+1}\!\left[\phi,\mathcal{P},\mathcal{O}\right]$ joining together the 3-point amplitude $\Amp_3[\phi\mathcal{P}\mathcal{Q}]$ (with $\mathcal{P}$ and $\mathcal{Q}$ two particles of different species) and the $n$-point amplitude $\Amp_n[\mathcal{Q},\mathcal{O}]$, where once more $\mathcal{O}$ denotes a set of arbitrary particles. We face, however, an immediate problem: when the soft limit is taken in this case, the particle $\mathcal{Q}$ in the propagator \textit{will not} be close to its mass shell, since by assumption its mass is different from the one of particle $\mathcal{P}$, and the factorization of Eq.\,\eqref{eq:soft_factorizarion} fails. Stated in other terms: given that $\Amp_{n+1}\!\left[\phi,\mathcal{P},\mathcal{O}\right]$ and $\Amp_n\!\left[\mathcal{Q},\mathcal{O}\right]$ contain, respectively, particle $\mathcal{P}$ and $\mathcal{Q}$ that are different, the two amplitudes cannot have the same kinematical configuration. This implies that the ALP soft factorization condition cannot hold in this case. 
To avoid this, we can work first in the high-energy limit, where all fermions are massless and Eq.\,\eqref{eq:soft_factor} holds again.
On the one hand, at leading order in $p_\phi$, the soft factor in Eq.\,\eqref{eq:soft_factor_toy} vanishes automatically as $\braket{pp}=\sbraket{pp}=0$ and regularity is trivially satisfied. On the other hand, the subleading terms that are constant in $p_\phi$ do not vanish and can be constrained in the collinear configuration. Considering that the angle between the 3-momenta is $\theta\ll 1$, we obtain for each helicity configuration (see Eq.\,\eqref{eq:collinear_product_massless})
\be\label{eq:S_psi_many}
\lim_{\theta\ll 1}\mathcal{S}_\psi \propto \frac{1}{\theta}.
\ee
Thus, using the regularity condition in the collinear limit, we arrive at the conclusion that the coefficients of the amplitude should vanish in the massless limit, {\it i.e.} they should be proportional to the mass of the fermions involved in the amplitude. The couplings, that are now matrices in the space of fermion species, can always be put into the form
\be\label{eq:gL/R}
g_{L,R} = \frac{i}{f}\left[M_\psi B_{L,R} - A_{L,R}M_\psi\right],
\ee
where $M_\psi$ is the fermion mass matrix and $A_{L,R},~B_{L,R}$ are arbitrary matrices, that can depend in a regular way on $M_\psi$ as well. If there is any massless fermion in the spectrum, or if some different species are degenerate in mass, we must treat them separately as the form of the amplitude is different, meaning that we can always take $M_\psi$ to be diagonal with positive non-degenerate entries. Since the parametrization above is redundant under the transformation $A_{L,R}\to A_{L,R} + M_\psi X_{L,R}$, $B_{L,R}\to B_{L,R} + X_{L,R}M_\psi$, with $X_{L,R}$ some matrix, we can always choose $A_{L,R}$ to be hermitian by choosing $X_{L,R}=M_\psi^{-1}A_{L,R}^\dagger$. Imposing $CPT$ and unitarity, which based on Eq.\,\eqref{eq:CPT_amp} amounts to $g_L^{\phantom{\dagger}}=g_R^\dagger$, we obtain that 
\be
g_L = \frac{i}{f}\left[M_\psi A_R - B_R^\dag M_\psi\right], ~~~ g_R = \frac{i}{f}\left[M_\psi B_{R} - A_{R}M_\psi\right],
\ee
{\it i.e.} the couplings $g_{L,R}$ depend on the hermitian matrix $A_R$ and on the complex matrix $B_R$. The matrix elements of $B_R$ can be simplified as follows. First of all, we observe that the real part of $(B_R)_{ii}$ can always be absorbed in $(A_R)_{ii}$, since the fact that $M_\psi$ is diagonal implies that $(B_R)_{ii}$ and $(A_R)_{ii}$ are not uniquely defined. Furthermore, since the diagonal couplings $(g_{L,R})_{ii}$ appear in the coupling between the ALP and two fermions of the same species, they must satisfy the soft factorization condition of Eq.\,\eqref{eq:soft_factorizarion}. As shown in Section\,\ref{subsec:fermions}, this forces their imaginary part to vanish. We thus conclude that the matrix $B_R$ can always be taken with vanishing diagonal elements. We still have enough freedom to significantly simplify the expression for $(B_R)_{ij}$, $i\neq j$. Indeed, applying the transformation $A_R \to A_R + M_\psi Y$, $B_R 
\to B_R + Y M_\psi$, with $Y$ satisfying $M_\psi Y = Y^\dag M_{\psi}$, the expressions for $g_{L,R}$ do not change and $A_R$ remains an hermitian matrix.\,\footnote{We thank Quentin Bonnefoy for a clarifying discussion about this point.} In order to satisfy $M_\psi Y = Y^\dag M_\psi$, the matrix elements of $Y$ must be of the form $Y_{ji} = M_i Y_{ij}^*/M_j$. The off-diagonal matrix elements $Y_{ij}$ ($i\neq j$) can then be chosen to impose the condition $(B_R)_{ji} = (B_R)_{ij}^*$ (it is sufficient to pick ${\rm Re}(Y_{ij}) = M_j({\rm Re}(B_{R,ji}) - {\rm Re}(B_{R,ij}))/(M_j^2 - M_i^2)$ and ${\rm Im} (Y_{ij}) = M_j ({\rm Im}(B_{R,ji}) + {\rm Im}(B_{R,ij}))/(M_i^2 - M_j^2)$, which is always allowed since the masses $M_i$ are assumed to be non-vanishing and non-degenerate), in such a way that the matrix $B_R$ can always be chosen to satisfy $B_R = B_R^\dag$ with vanishing diagonal elements. The couplings appearing in the amplitude are thus
\be\label{eq:coupling_toy_fermion_many}
g_{L} = \frac{i}{f}\left[M_\psi A_R - B_R M_\psi \right], ~~~ g_R = \frac{i}{f} \left[M_\psi B_{R} - A_{R}M_\psi\right],
\ee
with both $A_R$, $B_R$ hermitian. This expression agrees with what one would expect from a shift-symmetric coupling of one ALP to fermions\,\cite{Chala:2020wvs,Bauer:2020jbp,Bonnefoy:2022rik,DasBakshi:2023lca}, and precisely corresponds to the operator
$(1/f)(\del_\mu\phi)\bar \psi[C_V+C_A\gamma^5]\gamma^\mu \psi$, with 
$A_{R}=C_V+ C_A$ and $B_{R}=C_V- C_A$.
We observe that the diagonal elements of $C_V$ are not physical due to vector current conservation and can thus be redefined to obtain $(B_R)_{ii} = 0$, in agreement with what has been found above. In this case, that corresponds to the 1-family case, the dependency on $(C_V)_{ii}$ thus drops out and it follows from Eq.\,\eqref{eq:coupling_toy_fermion_many} that $g_L=-g_R$, in agreement with Eq.\,\eqref{eq:coupling_toy_fermion}.


We emphasise that the requirement of regularity of the soft factor $\mathcal{S}_\mathcal{P}$, combined with a well behaved massless limit of the low-energy (massive) amplitudes
and $CPT$ invariance plus unitarity, allowed us to fully determine the structure of the 3-point amplitudes up to some constants. In addition, they all agree with the amplitudes one would obtain by imposing shift-symmetry at the level of Lagrangian.

\subsection{Multiple soft limit}

We close this section by commenting on the case that Eq.\,\eqref{eq:soft_factorizarion} contains more than one ALP, that is, when multiple ALPs are being taken soft. In this situation, we can make recursive use of \textit{polology}  to obtain an expression similar to Eq.\,\eqref{eq:soft_limit_toy}, with, however, a different soft factor:
\begin{multline}\label{eq:multisoft}
    \lim_{p_{\phi_1},\cdots,p_{\phi_m}\to 0}  \Amp_{n+m}\!\left[\phi_1\cdots\phi_m\,\mathcal{P}^{I_1,\cdots ,I_{2s}}_{i,p},\cdots \right] = \\
    \lim_{p_{\phi_1},\cdots,p_{\phi_m}\to 0}\sum_{i=1}^n \Amp_n\!\left[\mathcal{P}^{K_1,\cdots, K_{2s}}_{i},\cdots\right] \times \left(\mathcal{S}^{(m)}_{\mathcal{P}_i}\right)_{K_1,\cdots, K_{2s}}^{I_1,\cdots, I_{2s}},
\end{multline}
where we have defined the $m$-th soft factor, $\mathcal{S}^{(m)}_{\mathcal{P}}$, associated to the soft emission of $m$ ALPs. Then, a natural extension of the ALP soft factorization condition\,\eqref{eq:soft_factor} is to impose that $\mathcal{S}^{(m)}_{\mathcal{P}}$ is regular for all $m\geq 1$. For $m=1$, we have explicitly computed the soft factor in Eq.\,\eqref{eq:soft_factor_toy}, that depended only on the 3-point amplitude $\Amp_3[\phi\mathcal{P\bar P}]$, and showed how this latter is constrained by the ALP soft factorization condition\,\eqref{eq:soft_factor}. For $m>1$, instead, the soft factor depends not only on the 3-point amplitude, but also on higher-point amplitudes involving more ALPs, $\Amp_4[\phi^2\mathcal{P\bar P}],\cdots,\Amp_{m+2}[\phi^m\mathcal{P\bar P}]$, since these are in general non-vanishing. To make the discussion more concrete, let us focus on the double soft limit, $m=2$ (see Fig.\,\ref{fig:multiple_soft}). 
\begin{figure}[t]
\centering
\begin{align*}
\lim_{p_{\phi_{1,2}}\to 0}
\adjustbox{valign=m}{\begin{tikzpicture}[line width=0.75]
\draw[fill=white] (0,0) circle (17pt) node[midway] {$\Amp_{n+2}$};
\draw (0.35,0.49) -- (1.2,1) node[right] {$\mathcal{P}_{i,p}$};
\draw[dashed] (0.35,-0.49) -- (1.2,-1) node[right] {$\phi_1$};
\draw (-0.35,0.49) -- (-1.2,1);
\draw[dashed] (-0.35,-0.49) -- (-1.2,-1)node[left] {$\phi_2$};
\filldraw[black] (-1.3,0) circle (0.75pt);
\filldraw[black] (-1.2,0.3) circle (0.75pt);
\filldraw[black] (-1.2,-0.3) circle (0.75pt);
\filldraw[black] (1.3,0) circle (0.75pt);
\filldraw[black] (1.2,0.3) circle (0.75pt);
\filldraw[black] (1.2,-0.3) circle (0.75pt);
\end{tikzpicture}}
& =\lim_{p_{\phi_{1,2}}\to 0}\left\{
\adjustbox{valign=m}{\begin{tikzpicture}[line width=0.75]
\draw[fill=white] (0,0) circle (16pt) node[midway] {$\Amp_{n}$};
\draw (-1+0.55,0.35) -- (-1.,0.7);
\draw (-1+0.55,-0.35) -- (-1.,-0.7);
\filldraw[black] (-1.2+0.1,0) circle (0.75pt);
\filldraw[black] (-1.1+0.1,0.25) circle (0.75pt);
\filldraw[black] (-1.1+0.1,-0.25) circle (0.75pt);
\draw (0.55,0) -- (2-0.75,0);
\draw[fill=white] (2.55-0.75,0) circle (16pt) node(2,55,0) {$\Amp_{3}$};
\draw[dashed] (2.55-0.75,0.55) -- (2.55-0.75,1.15)node[above]{$\phi_2$};
\draw (2.35,0) -- (3.05,0);
\draw[fill=white] (2.55+1,0) circle (16pt) node(2,55,0) {$\Amp_{3}$};
\draw (1-0.55+2.55+1,0.35) -- (1.3+1.55+1.5,0.7) node[right] {$\mathcal{P}_{i,p}$};
\draw[dashed] (1-0.55+2.55+1,-0.35) -- (1.3+2.55+0.5,-0.7) node[right] {$\phi_1$};
\end{tikzpicture}} \right.\\
& ~~~~~~~~~~~~~~~~~~~~~~~~~~~~~+ (\phi_1 \leftrightarrow \phi_2) \\
& ~~~~~~~~~~~~~~~ \left.+~~ \adjustbox{valign=m}{\begin{tikzpicture}[line width=0.75]
\draw[fill=white] (0,0) circle (16pt) node[midway] {$\Amp_{n}$};
\draw (-1+0.55,0.35) -- (-1.,0.7);
\draw (-1+0.55,-0.35) -- (-1.,-0.7);
\filldraw[black] (-1.2+0.1,0) circle (0.75pt);
\filldraw[black] (-1.1+0.1,0.25) circle (0.75pt);
\filldraw[black] (-1.1+0.1,-0.25) circle (0.75pt);
\draw (0.55,0) -- (2-0.4,0) ;
\draw[fill=white] (2.55-0.4,0) circle (16pt) node(2,55,0) {$\Amp_{4}$};
\draw (2.55+0.55-0.4,0) -- (1.3+2.55-0.6,0) node[right] {$\mathcal{P}_{i,p}$};
\draw[dashed] (1-0.55+2.55-0.4,0.35) -- (1.3+2.55-0.6,0.8) node[right] {$\phi_2$};
\draw[dashed] (1-0.55+2.55-0.4,-0.35) -- (1.3+2.55-0.6,-0.8) node[right] {$\phi_1$};
\end{tikzpicture}} + \dots \right\}
\end{align*}
\caption{Schematic diagrammatic representation of Eq.\,\eqref{eq:multisoft} for an amplitude with two ALPs ($m=2$). The first line represents the insertion of two 3-point amplitudes, the second one is the Bose exchange of the two ALPs in the double $\Amp_3$ insertion, while the last line represents the insertion of the 4-point amplitude with two ALPs whose coefficients need to be determined. The dots denote other factorization channels that are automatically regular in the double soft limit.}
\label{fig:multiple_soft}
\end{figure}
The soft factor in this case, $\mathcal{S}^{(2)}_{\mathcal{P}}$, contains a piece that depends on two insertions of the 3-point amplitude\,\eqref{eq:3_pointamp_toy}, and a second piece that depends on $\Amp_4[\phi^2\mathcal{P\bar P}]$. Since the 3-point amplitude is fixed by the ALP soft factorization condition\,\eqref{eq:soft_factor}, we would expect that imposing the regularity of the double soft limit would allows us to constraint the 4-point amplitude, $\Amp_4[\phi^2\mathcal{P\bar P}]$. Already at this level, however, things are not so simple. Doing the explicit computation, we find that $\mathcal{S}_{\mathcal{P}}^{(2)}$ has a term that can be canceled by a suitable choice of the coefficients appearing in $\Amp_4[\phi^2\mathcal{P\bar P}]$, but other divergent terms remain whose spinor helicity structures are not obviously connected to any tree level amplitude. We suspect that the solution to this problem lies in loop corrections, as it happens in the discussion of multiple soft limits of photons\,\cite{Weinberg:1996kr}. In that case, radiative corrections are instrumental in guaranteeing that the overall amplitude is finite and the same may happen in our case. Given the complexity of general loop computation in the on-shell context, we defer the detailed analysis of this class of amplitudes to future work.

\textcolor{white}{a}

\section{ALP interactions with the SM particles}\label{sec:SM}

Having constructed the 3-point amplitudes that describe the ALP interactions to matter, we now discuss how to extend our results to the case of the SM. There are two different regimes for the SM amplitudes. One is when (some of) the particles are massive and the amplitudes are invariant under the symmetry group $SU(3)_c\times U(1)_\text{EM}$, corresponding in the usual QFT language to the broken phase of electroweak symmetry. These amplitudes are the relevant ones at low-energies, that is, when the typical energy scale is much smaller than the scale of electroweak symmetry $v$ (to be defined below). For massive amplitudes in this low-energy regime, apart for some overall group theoretical factors, the results of Section\,\ref{sec:toy} can be directly applied. For energies above the scale $v$, \textit{i.e.} at high-energies, we take the particle content to be massless and amplitudes to be invariant under $\mathcal{G}_\text{SM}=SU(3)_c\times SU(2)_L\times U(1)_Y$, thus corresponding to the unbroken electroweak phase. Explicit expressions for these amplitudes involving also the ALP will be given below. The link between low- and high-energy amplitudes will allow us to find additional properties of the coefficients $C_{\phi\psi\psi}$, $C_{\phi VV}$ and $g_{L,R}$ introduced in the previous section, once the connection with the coefficients of the high-energy amplitude is made explicit. In practice, we will take the massless limit of the massive amplitudes (as defined in Appendix\,\ref{app:conventions}) and equate them to the corresponding $\mathcal{G}_\text{SM}$-invariant massless amplitudes, under the assumption that the $m\to0$ limit is smooth and continuous.

The remainder of this section will be devoted to the study of the high-energy limit applied to the amplitudes involving one ALP and different particles. The case of the coupling between one ALP and two Higgs doublets is identical to the one already discussed in Eq.\,\eqref{eq:S_s}, so we can skip directly to the coupling between the ALP and particles of higher spin. For spin 1 particles (analyzed in Section\,\ref{sec:phiVV_SM}) the high-energy limit of Eq.\,\eqref{eq:coupling_toy_vector} will be straightforward and will allow us to relate the coefficients appearing in the amplitudes involving $\phi WW$, $\phi ZZ$, $\phi \gamma\gamma$ and $\phi \gamma Z$ to the coefficients appearing in the amplitudes involving $\phi BB$ and $\phi W^a W^a$ (with $a$ the $SU(2)_L$ adjoint index). The case of the amplitude $\phi G^A G^A$ (where $G$ is a gluon and $A$ is the index of the $SU(3)_c$ adjoint) will be more difficult, since this amplitude cannot be linked to any of the massive amplitudes derived in Section\,\ref{sec:toy}.
In this case, we will need to resort to a one-loop computation to recover the same structure of the amplitude that emerges for the other spin 1 bosons.
The case of spin 1/2 particles (discussed in Section\,\ref{sec:phiff_SM}) will be even trickier, since $\mathcal{G}_\text{SM}$-invariance will force us to connect low-energy 3-point amplitudes to high-energy 4-point amplitudes involving one additional Higgs doublet. As we are going to see, in this case electroweak symmetry breaking (as discussed in Ref.\,\cite{Balkin:2021dko}) will be essential to meaningfully connect low- and high-energy coefficients and will force us to introduce a ``ALP-Higgs obstruction'' condition to complement the ALP soft factorization condition of Eq.\,\eqref{eq:soft_factor}. In this section we will still be concerned with amplitudes that involve only one ALP that are suppressed by $f^{-1}$. We will discuss in Section\,\ref{sec:higher} how to compute amplitudes suppressed by more powers of $f$ and we will comment on the limitations of our approach.



\subsection{$\phi VV$ amplitudes}\label{sec:phiVV_SM}

The massless 3-point amplitude involving one ALP and two spin-1 particles can be built using little group covariance and $\mathcal{G}_\text{SM}$-invariance. The result is
\al{\label{eq:ALP_2ewvectors_SM}
\Amp_3\!\left[\phi B_1^-B_2^-\right] = \frac{g_B^-}{f}\braket{12}^2,&~\Amp_3\!\left[\phi B_1^+B_2^+\right] = \frac{g_B^+}{f}\sbraket{12}^2,\\
\Amp_3\!\left[\phi W_1^-W_2^-\right] = \frac{g_W^-}{f}\braket{12}^2,&~\Amp_3\!\left[\phi W_1^+W_2^+\right] = \frac{g_W^+}{f}\sbraket{12}^2,
}
where, for simplicity, we have ignored $\mathcal{G}_\text{SM}$ indices. The $\pm$ superscripts show explicitly the helicities of the spin 1 particles.\footnote{We do not consider amplitudes with one positive and one negative helicity states, as these amplitudes have a more intricate analytical structure and do not necessarily vanish as the momenta are taken real\,\cite{Benincasa:2007xk,Benincasa:2011pg,Benincasa:2011kn}.} We recall that massless 3-point amplitudes are non-vanishing only when considering complex momenta. A priori, little group covariance allows the a priori complex coefficients $g_{B,W}^\pm$ to be arbitrary and unrelated one with the other. As in Section\,\ref{sec:toy}, we now want to find whether any correlation exists between these coefficients and how to connect such coefficients with those appearing in the amplitudes involving physical states. Let us start with the 3-point amplitudes involving one ALP and the massive spin-1 particles, $\Amp_3[\phi W_1^{(+)} W_2^{(-)}]$ (where the superscripts now denote the electric charge of the massive $W$ bosons, not to be confused with the helicities that appear in Eq.\,\eqref{eq:ALP_2ewvectors_SM} for the massless $W$ bosons in the unbroken phase) and $\Amp_3[\phi Z_1 Z_2]$. As we saw in Section\,\ref{sec:toy}, in this case the ALP soft factorization condition \eqref{eq:soft_factor} holds and the amplitudes are those of Eq.\,\eqref{eq:coupling_toy_vector}, with real coefficients $C_{\phi WW}$ and $C_{\phi ZZ}$. Taking the high energy limit of these massive amplitudes and using spontaneous symmetry breaking as implemented in the on-shell context\,\cite{Bachu:2019ehv,Bachu:2023fjn}, we obtain the following correspondence between $C_{\phi WW, \phi ZZ}$ and $g_{B,W}^\pm$:\,\footnote{As discussed in Appendix\,\ref{app:conventions}, in the cases considered in this paper, when taking the high energy limit of the massive spinor helicity structure $\langle \bm{12} \rangle^2$ we simply obtain the massless structure $\langle 12\rangle^2$, which selects negative helicity for the spin-1 particles involved. In an analogous way, the massive structure $[\bm{12}]^2$ corresponds to the massless $[12]^2$ and positive helicity vectors. From this correspondence, the result in Eq.\,\eqref{eq:CphiZZ_CphiWW} immediately follows.}
\be\label{eq:CphiZZ_CphiWW}
g_{B,W}^\mp = - g_{B,W}^\pm , \quad i C_{\phi ZZ} = \sin^2(\theta_W) g_B^- + \cos^2(\theta_W) g_W^-, \quad iC_{\phi WW} = g_W^-,
\ee
where $\theta_W$ is the weak angle. We thus see that, although little group covariance does not correlate $g_{B,W}^\pm$ one with the other, using the high energy limit and electroweak symmetry breaking we are not only able to (anti) correlate $g_{B,W}^+$ with $g_{B,W}^-$, but we are also able to express the coefficients that appear in the massive amplitudes, $C_{\phi ZZ}$ and $C_{\phi WW}$, with the high energy coefficients $g_{B,W}^\pm$. The remaining amplitudes involving the physical spin-1 states (other than the gluons, to which we will come back later in this section) are $\Amp_3[\phi \gamma_1^\pm \gamma_2^\pm]$ and $\Amp_3[\phi Z_1 \gamma_2^\pm]$, where $\gamma^\pm$ denotes a photon with helicity $\pm 1$. The form of these amplitudes is completely fixed by little group covariance to be
\al{
\Amp_3\!\left[\phi \gamma_1^-\gamma_2^-\right] = \frac{g_{\gamma\gamma}^-}{f}\braket{12}^2,&~\Amp_3\!\left[\phi \gamma_1^+\gamma_2^+\right] = \frac{g_{\gamma\gamma}^+}{f}\sbraket{12}^2,\\
\Amp_3\!\left[\phi Z_1\gamma_2^-\right] = \frac{g_{Z\gamma}^-}{f}\braket{\bm{1}2}^2,&~\Amp_3\!\left[\phi Z_1\gamma_2^+\right] = \frac{g_{Z\gamma}^+}{f}\sbraket{\bm{1}2}^2.
}
Once more, the coefficients appearing are, a priori, unrelated one with the other. We can, however, use once more electroweak symmetry breaking as implemented in Ref.\,\cite{Bachu:2019ehv} to obtain
\be
i C_{\phi\gamma\gamma} = g_{\gamma\gamma}^- = - g_{\gamma\gamma}^+, ~~~ i C_{\phi Z\gamma} = g_{Z\gamma}^- = - g_{Z\gamma}^+,
\ee
where
\be\label{eq:Cgammagamma}
i C_{\phi\gamma\gamma} = \sin^2(\theta_W) g_W^- + \cos^2(\theta_W) g_B^-, ~~~ i C_{\phi Z\gamma} = 2\cos(\theta_W) \sin(\theta_W) \left[ g_W^- - g_B^-\right].
\ee
We thus obtain the non-trivial result that all the coefficients that appear in the 3-point amplitudes between physical states are correlated one with the other through their dependence on the high-energy coefficients $g_B^\pm$ and $g_W^\pm$. In terms of degrees of freedom, we have started from four a priori complex parameters ($g_{W,B}^\pm$) and reduced them to only 2 real ones.

Gluons, on the other hand, cannot be related to other gauge bosons via symmetry breaking, nor are they massive at low energy. At all energies at which perturbation theory is valid, the coupling between the ALP and gluons will be
\be
\label{eq:ALP_2gluons_SM}
\Amp_3\!\left[\phi G_1^-G_2^-\right] = \frac{g_G^-}{f}\braket{12}^2,~\Amp_3\!\left[\phi G_1^+G_2^+\right] = \frac{g_G^+}{f}\sbraket{12}^2,
\ee
as usual omitting $\mathcal{G}_\text{SM}$ structures. Following our discussion above, we would expect the coefficients $g_G^-$ and $g_G^+$ to be related by $g_G^- = - g_G^+$ as in the case of the photons, but in this case we cannot relate the amplitudes under consideration with massive amplitudes (gluons are never massive) nor we can use electroweak symmetry breaking as we did for photon (color symmetry is never broken). The way we found to extract some information out of these amplitudes is to go to the 1-loop level and see the interplay between $\Amp_3[\phi G^\pm G^\pm]$ and 3-point amplitudes involving fermions. More precisely, we compute the anomalous dimension of the coefficient $C_\psi$ appearing in the amplitude between $\phi$ and two fermions $\psi$ (the coupling will be defined in Eq.\,\eqref{eq:ALP_2fermions_SM}) induced by Eq.\,\eqref{eq:ALP_2gluons_SM}. The coefficient $C_\psi$ cannot be arbitrary, but in order to satisfy the ALP soft factorization condition (or, more precisely, to satisfy the ``ALP-Higgs obstruction condition'' that we will introduce in Section\,\ref{sec:phiff_SM} and that will be added to the ALP soft factorization condition), it will have to satisfy a specific constraint, shown in Eq.\,\eqref{eq:coupling_SM_fermion}. Our 1-loop computation (shown in Appendix~\ref{app:running}) shows that $C_\psi$ can satisfy Eq.\,\eqref{eq:coupling_SM_fermion} only if
\be\label{eq:Cgluglu}
g_G^-=-g_G^+=iC_{\phi GG},
\ee
with $C_{\phi GG}$ real. We thus conclude that the coefficients appearing in the amplitudes with gluons do indeed satisfy the same relation we found in the case of photons. It is worth stressing, however, that the reasoning above makes the amplitudes in Eq.\,\eqref{eq:ALP_2gluons_SM} qualitatively different from the others, as we were only able to arrive at the constraints above by using other amplitudes.


Summarizing, the amplitudes in Eqs.\,\eqref{eq:ALP_2ewvectors_SM} and \eqref{eq:ALP_2gluons_SM} are constrained to be of the form
\be\label{eq:ALP_2vectors_SM_final}
\Amp_3\!\left[\phi V^-_1 V_2^-\right]=\frac{iC_{\phi VV}}{f}\braket{12}^2,\quad \Amp_3\!\left[\phi V^+_1 V_2^+\right]=-\frac{iC_{\phi VV}}{f}\sbraket{12}^2,
\ee
with $V=B,W$ or $G$, and $C_{\phi VV}$ real. 
For the phenomenology of ALPs, the most important couplings are the one with photons and gluons, which we define at the Lagrangian level by $\La = \frac{C_{\phi\gamma\gamma}}{2f} \phi F_{\mu\nu} \tilde{F}^{\mu\nu} + \frac{C_{\phi GG}}{2f} \phi G^A_{\mu\nu} \tilde{G}^{A\mu\nu}$, with the couplings $C_{\phi\gamma\gamma}$ and $C_{\phi GG}$ given in terms of the high-energy couplings by Eqs.\,\eqref{eq:Cgammagamma} and \eqref{eq:Cgluglu}.


\subsection{Fermions and EWSB}\label{sec:phiff_SM}

The case of fermions is more subtle and, as we already mentioned at the beginning of this section, to properly address it we will need to add a condition to Eq.\,\eqref{eq:soft_factor}. The problem stems from the fact that, unlike what happens in Eq.\,\eqref{eq:ALP_2fermions_toy}, we cannot build an amplitude with one ALP and two fermions due to $\mathcal{G}_\text{SM}$-invariance and fermion helicities. 
Nevertheless, it is possible to add a Higgs doublet and write
\be\label{eq:ALP_2fermions_SM}
\Amp_4\!\left[\phi \psi_{L1}^-\bar \psi_{R2}^- \bar H\right] = \frac{\bar C_\psi}{f}\braket{12},~~ \Amp_4\!\left[\phi \bar \psi_{L1}^+ \psi_{R2}^+ H\right] = \frac{C_\psi}{f}\sbraket{12},
\ee
where $\psi_{L,R}$ denote the chiral fermions of the SM, while $H$ is the Higgs doublet and $\bar H$ denotes the anti-Higgs (in the case of couplings to up-quarks one should swap $H\leftrightarrow\bar H$).  For simplicity, we again suppress all indices and tensor structures related to $\mathcal{G}_\text{SM}$. Also, it is important to notice that the dimensionless couplings $C_\psi$, $\bar C_\psi$ are matrices in fermion flavor space and the spinor structures $\braket{12}$ and $\sbraket{12}$ are flavor-independent since all fermions are massless. This also implies that we have the freedom to redefine $C_\psi$ and $\bar C_\psi$, because in the massless limit they can be seen as tensors of the flavor group $U(3)_{\psi_L}\times U(3)_{\psi_R}$\,\cite{Alves:2021rjc}. At tree-level, $CPT$ and unitarity enforces that $\bar C_\psi^{\phantom{\dagger}} = C_\psi^\dagger$, that follows from Eq.\,\eqref{eq:CPT_amp}.

The fact that the amplitude in Eq.\,\eqref{eq:ALP_2fermions_SM} is now a 4-point amplitude brings two changes to the analysis. First, the coefficients $C_\psi$ can now depend on the kinematics through kinematical invariants, which we will always assume to be regular, \textit{i.e.} given by a power expansion. Second, the reasoning that led us to the ALP soft factorization condition \eqref{eq:soft_factor} fails, because taking $p_\phi\to0$ does not guarantee that the particle exchanged in the propagator goes on-shell and that the amplitude factorizes. The key to understand the soft limit for these amplitudes lies in the Brout--Englert--Higgs mechanism. In general, to connect the massless (high-energy) amplitudes with the massive (low-energy) amplitudes without the Higgs, one uses
\be\label{eq:higgsing}
\lim_{p_H\to 0}\Amp\!\left[H,\cdots\right] = \lim_{\text{H.E.}}\frac{1}{v}\Amp\!\left[\cdots\right],
\ee
where H.E. stands for the high-energy limit of the amplitude (see Appendix\,\ref{app:conventions}) and $v$ is the scale at which the Higgs becomes non-dynamical (``frozen'' in the language of Ref.\,\cite{Balkin:2021dko}), which is introduced by dimensional analysis and represents the scale associated with electroweak symmetry. The limit $p_H\to 0$ is to be understood also as the limit in which the Higgs becomes non-dynamical, which amounts to removing it from the amplitude, and the right hand side is to be taken as the high-energy (massless) limit of the corresponding massive amplitude. What is stated in Eq.\,\eqref{eq:higgsing} is nothing but the UV/IR compatibility for amplitudes involving the Higgs.

In short, we learn from the discussion above that the ALP soft factorization condition \eqref{eq:soft_factor} is not directly applicable to amplitudes involving Higgs. Therefore, we are led to impose an extra condition for such amplitudes:
\begin{GrayBox}
\textbf{ALP-Higgs obstruction}
\vspace{0.5em}
\textit{If a low-energy 3-point amplitude $\Amp_3$ involving one massless ALP cannot be associated to a non-vanishing gauge invariant 3-point amplitude in the UV, but only to a 4-point one with an extra Higgs, then 
we impose the following condition:}
\be\label{eq:Higgs_obstruction}
\lim_{p_\phi,p_H\to 0}\Amp_4[\phi\, H \cdots] = \lim_{p_\phi\to 0}\lim_{\text{H.E.}} \frac{1}{v}\Amp_3[\phi\cdots].
\ee
\end{GrayBox}
The prescription above guarantees that the result we obtained previously in Eq.\,\eqref{eq:coupling_toy_fermion_many} holds, as the double soft-limit $p_\phi,p_H\to 0$ assures that the factorization in Eq.\,\eqref{eq:soft_factorizarion} takes place.\,\footnote{
We stress that this condition is reminiscent of the {\it Higgs low-energy theorems}~\cite{Ellis:1975ap,Shifman:1979eb,Vainshtein:1980ea}. However, these theorems, unlike in our case, are typically employed in the broken phase of electroweak symmetry. And they are used to obtain contributions from loop amplitudes coming from heavy particles that couple to the Higgs, which is not what is stated in Eq.\,\eqref{eq:Higgs_obstruction}.
Nonetheless, it would be interesting to derive an on-shell perspective of these theorems, which is still missing as far as we know.
}
Therefore, from Eq.\,\eqref{eq:coupling_toy_fermion_many}, after applying Eq.\,\eqref{eq:higgsing}  to \eqref{eq:ALP_2fermions_SM}, we obtain
\be\label{eq:coupling_SM_fermion}
C_\psi\left(p_H=0\right) = i\left(Y_\psi \tilde B_R - \tilde A_R Y_\psi\right),
\ee
where $Y_\psi$ are the diagonal Yukawa couplings, $\tilde B_{R}^{\phantom{\dagger}}= \lim_{\text{H.E.}} U_{R}^{\phantom{\dagger}}B_{R}^{\phantom{\dagger}} U_{R}^\dagger$ and\linebreak $\tilde A_{R}^{\phantom{\dagger}}= \lim_{\text{H.E.}} U_{L}^{\phantom{\dagger}}A_{R}^{\phantom{\dagger}} U_{L}^\dagger$ (with $A_{R}^{\phantom{\dagger}},B_{R}^{\phantom{\dagger}}$ the original couplings appearing in Eq.\,\eqref{eq:coupling_toy_fermion_many})
The equation above needs clarification. The first point regards the appearance of flavor transformations. As mentioned previously, when the fermions are massless we are free to perform $U(3)_{\psi_L} \times U(3)_{\psi_R}$ flavor transformations, while this freedom is lost when they become massive. Hence, when connecting the massive and massless regimes,
amplitudes will in general agree up to a flavor transformation\,\cite{Alves:2021rjc}. Secondly, the Yukawa matrices can be expressed here as $vY_\psi^{\phantom{\dagger}} = U_L^{\phantom{\dagger}} M_\psi^{\phantom{\dagger}} U_R^\dagger$, where the relation to the mass from an on-shell perspective was shown in Refs.\,\cite{Balkin:2021dko,Bachu:2023fjn} and we provide an alternative derivation in Appendix~\ref{app:Yukawa-mass}. Moreover, from the two previous points we can derive the expressions for $\tilde A_{R},\tilde B_R$, recalling that the matrices $A_{R},B_R$
can depend on higher powers of the fermion masses, such that the high-energy limit selects only the mass-independent term.
At zero Higgs momentum, that is, for constant coupling, Eq.\,\eqref{eq:coupling_SM_fermion} is identical to what one would expect from ALPs coupled to fermions in a shift-symmetric manner\,\cite{Chala:2020wvs,Bauer:2020jbp,Bonnefoy:2022rik,DasBakshi:2023lca}.

Before moving on to other amplitudes, it is worth to explore in more detail some features of the amplitude $\Amp_3\!\left[\phi \psi\bar\psi\right]$ and its relation to the ALP-Higgs obstruction.
At face value, 
the high-energy limit of this amplitude with couplings given in\linebreak Eq.\,\eqref{eq:coupling_toy_fermion_many} leads to a vanishing result, since $C_\psi\sim M_\psi$.
This naive result is inconsistent with $\mathcal{G}_\text{SM}$-invariance and, for this reason, we had to promote the UV amplitude to a 4-point amplitude with an additional Higgs. One could, however, imagine an alternative route. Using the Weyl equations in Eq.\,\eqref{eq:Weyl}, it is possible to trade the mass factor for a momentum insertion in the spinor structures, {\it e.g.} $M_\psi\braket{\bm{12}}\sim \asbraket{\bm{1}}{p_\phi}{\bm{2}}$ for the case of angle brackets, with an analogous identity holding for square brackets. This form highlights that the amplitude does not vanish in the high-energy limit and should be directly matched into $\mathcal{G}_\text{SM}$-invariant 3-point amplitudes $\Amp_3\!\left[\phi\psi_{L,R}\bar \psi_{L,R}\right]$. But these amplitudes nonetheless vanish identically due to the 3-particle kinematics, leaving us with the 4-point amplitude already discussed as the unique route to match the massive amplitude in the UV.

Another interesting feature of the coefficient in Eq.\,\eqref{eq:coupling_SM_fermion} is its connection with the invariants that parameterise the breaking of shift-symmetry defined in Ref.\,\cite{Bonnefoy:2022rik}. In this reference, a total of 3 and 10 invariants were constructed in the lepton and quark sector, respectively, using the mathematical properties of $C_\psi(p_H=0)$. From our perspective, the only way to break the coupling structure of Eq.\,\eqref{eq:coupling_SM_fermion} is to violate the scaling with the Yukawa matrices (which correspond to a breaking of the dependence on the particle mass in Eq.\,\eqref{eq:gL/R}), as the phase of the amplitude and the hermiticity of $A_R,B_R$
are fixed by $CPT$ and unitarity. We thus conclude that the maximum number of independent parameters that can break the shift-symmetry (in our language, that gives a singular massless/high energy limit) amounts to the number of independent parameters in the Yukawa matrices, {\it i.e.} the number of parameters left after all possible flavor transformations are applied. For leptons, taking massless neutrinos, this implies 3 independent parameters; for quarks (with two Yukawa matrices that are correlated in the UV because we have only one left handed doublet), the number increases to 10 (6 masses plus the 4 parameters of the Cabibbo--Kobayashi--Maskawa matrix\,\footnote{The emergence of the Cabibbo--Kobayashi--Maskawa and the Pontecorvo--Maki--Nakagawa--Sakata matrices from the on-shell perspective was studied in Ref.\,\cite{Alves:2021rjc}.}). Hence, in the general case in which the Yukawas have the maximal amount of parameters, this counting exactly corresponds to the one found in Ref.\,\cite{Bonnefoy:2022rik}. Since, however, the invariants are constructed based on the exact form of the flavor group, we cannot conclude that by reducing the physical parameters of the Yukawas will necessarily reduce the amount of non-vanishing invariants. In general, their number will only decrease once the flavor group is enlarged.

\vspace{1em}
\section{Constructing higher-point functions}\label{sec:higher}

So far we dealt with 3-point functions, {\it i.e.} amplitudes with couplings proportional to $1/f$ that, at leading order, can be matched onto dimension $d=5$ operators. We now proceed to extend our discussion to amplitudes of higher-order and build explicitly the amplitude basis up to $1/f^4$. In the literature, only $d=6$ (for instance in Refs.\,\cite{Bauer:2017ris, Brivio:2021fog}) and a couple of $d=7$ operators\,\cite{Bauer:2017ris,Bauer:2018uxu,Bauer:2016zfj,Bauer:2016ydr} were considered and studied at the phenomenological level. Our present work provides the complete amplitudes, and a corresponding operator basis, up to $d=8$ consistent with the ALP properties. Our results agree with what was recently found in the literature, in particular Refs.\,\cite{Song:2023lxf,Song:2023jqm}, that use a different on-shell approach, and Ref.\,\cite{Grojean:2023tsd} that employs Hilbert series techniques.

\subsection{Contact amplitudes}

The higher-point amplitudes we are interested in are contact ones, \textit{i.e.} amplitudes that are regular in the kinematical invariants, and necessarily involve 4 or more particles. These amplitudes, from now on denoted as $\Amp^\text{ct}$, are part of the physical amplitudes, that is simply the sum of contact and factorizable terms:
\be
\Amp_\text{phys} = \Amp^\text{ct} + \Amp^\text{fact}.
\ee
The factorizable terms are those that can be obtained joining together lower point amplitudes, using \textit{polology} to guarantee the correct pole structure\,\cite{Elvang:2013cua,Arkani-Hamed:2017jhn}. The strategy is similar to the one we outlined in Section\,\ref{sec:toy} when taking the soft limit: the structure of $\Amp^\text{fact}$ should be such that, when it is possible to exchange an intermediate particle, in the limit in which the intermediate particle goes on-shell the factorizable amplitude factorizes, \textit{i.e.} it can be written as the product of the lower point amplitudes times the propagator of the intermediate particle. Schematically, we thus have $\Amp^\text{fact}_{n_1 + n_2 -2} = \Amp^{(1)}_{n_1} \Amp^{(2)}_{n_2}/(p^2-m^2)$, where $\Amp^{(1,2)}_{n_{1,2}}$ are the two lower point amplitudes (a $n_1$-point and a $n_2$-point amplitude, respectively) and the intermediate particle has momentum $p$. For simplicity, we are leaving implicit the Levi-Civita tensors that ensure the correct transformation under little group transformations and that depend on the spin of the exchanged particles. Explicit examples of how the factorizable amplitudes can be constructed can be found in Ref.\,\cite{Arkani-Hamed:2017jhn}.

On the other hand, given that $\Amp^\text{ct}$ are contact amplitudes (\textit{i.e.} they do not contain propagators), they are by definition regular as $p_\phi\to 0$, so we cannot extract any information from them using the ALP soft factorization condition \eqref{eq:soft_factor} alone. In order to make progress, we need an additional physical constraint on them. We impose:

\begin{GrayBox}
\textbf{ALP soft contact condition}

\vspace{0.5em}

\textit{If $\Amp_n^\mathrm{ct}\!\left[ \cdots\right]$ is a contact amplitude with $n\geq 4$ involving at least one massless ALP, then
\be\label{eq:Adler}
\lim_{p_\phi\to 0}\Amp_n^\mathrm{ct}\!\left[ \cdots\right] = 0,
\ee
for each ALP $\phi$ present in the amplitude.} 
\end{GrayBox}
If we look back at the results from Section~\ref{sec:toy}, the only reason that the soft-limit did not give a zero was because of the $p_\phi$-independent terms, that only appeared because of the singular propagator in Eq.\,\eqref{eq:soft_factor_toy}. Since, now, we do not have such singular terms as we are looking directly at the contact amplitudes, the natural extension of the previous requirement is to have the amplitudes to vanish in the soft-limit. The only exceptions we found to the condition above are the $\Amp_4\!\left[\phi\psi \bar\psi H\right]$ amplitudes \eqref{eq:ALP_2fermions_SM}, that have a constant coefficient in Eq.\,\eqref{eq:coupling_SM_fermion}. However, as discussed in Section~\ref{sec:SM}, these amplitudes must be treated according to the ALP-Higgs obstruction \eqref{eq:Higgs_obstruction}.

We stress that the regularity conditions of Eqs.\,\eqref{eq:soft_factor} and \eqref{eq:Adler} are, in most cases, equivalent to the Adler's zero condition\,\cite{Weinberg:1996kr}. In Quantum Field Theoretical language the Adler's zero condition states that, in the soft limit $p_\phi \to 0$, the soft factor is regular when $\phi$ is emitted from some external leg and vanishes when $\phi$ is emitted from the interior of the diagram. The only exceptions are, as explained above, those amplitudes involving the Higgs doublet that match into 3-point amplitudes at low energy.

\begin{table}[t]
\centering
\renewcommand{\arraystretch}{1.5}
\begin{tabular}{ccc}
\multicolumn{3}{c}{Dimension 6} \\ \hline\hline
\multicolumn{1}{c|}{Particle content\,} & \multicolumn{1}{c|}{$\Amp^\text{ct}\times f^2$} & $\mathcal{O}\times f^2$ \\ \hline
\multicolumn{1}{c|}{$\phi^2H \bar H$} & \multicolumn{1}{c|}{$~p_{\phi_1}\cdot p_{\phi_2}~$} & $~(\del_\mu\phi)(\del^\mu\phi)|H|^2~$
\end{tabular}
\renewcommand{\arraystretch}{1}
\caption{Contact amplitudes suppressed by $1/f^2$ and corresponding $d=6$ operators. All the amplitudes are stripped of overall coefficients.}\label{tab:dim6}
\end{table}

\begin{table}[t]
\centering
\renewcommand{\arraystretch}{1.5}
\begin{tabular}{ccc}
\multicolumn{3}{c}{Dimension 7} \\ \hline\hline
\multicolumn{1}{c|}{Particle content} & \multicolumn{1}{c|}{$\Amp^\text{ct} \times f^3$} & $\mathcal{O}\times f^3$ \\ \hline
\multicolumn{1}{c|}{\multirow{2}{*}{$\phi \psi_1\bar\psi_2 H$}} & \multicolumn{1}{c|}{$~(p_{\phi}\cdot p_{1})\braket{12}~$} & $~(\del_\mu\phi)\bar\psi H (D^\mu\psi)~$ \\ \cline{2-3} 
\multicolumn{1}{c|}{} & \multicolumn{1}{c|}{\textbf{$~(p_{\phi}\cdot p_{2})\braket{12}~$}} & $~(\del_\mu\phi)(D^\mu\bar\psi) H \psi~$ \\ \hline
\multicolumn{1}{c|}{\multirow{4}{*}{$\phi \psi_1 \bar \psi_2 V_3$}} & \multicolumn{1}{c|}{\multirow{4}{*}{$\braket{13}\asbraket{3}{p_\phi}{2},\braket{23}\asbraket{3}{p_\phi}{1}$}} & $(\del^\mu\phi)\bar \psi \gamma^\nu \psi V_{\mu\nu}$ \\
\multicolumn{1}{c|}{} & \multicolumn{1}{c|}{} & $(\del^\mu\phi)\bar \psi \gamma^\nu \psi \tilde V_{\mu\nu}$ \\
\multicolumn{1}{c|}{} & \multicolumn{1}{c|}{} & $(\del^\mu\phi)\bar \psi \gamma^\nu T^A \psi V^A_{\mu\nu}$ \\
\multicolumn{1}{c|}{} & \multicolumn{1}{c|}{} & $(\del^\mu\phi)\bar \psi \gamma^\nu T^A\psi \tilde V^A_{\mu\nu}$ \\ \hline
\multicolumn{1}{c|}{\multirow{4}{*}{$\phi H_1 \bar H_2 V_3$}} & \multicolumn{1}{c|}{\multirow{4}{*}{$\bra{3}p_\phi(p_1-p_2)\ket{3}$}} & $(\del_\mu\phi)(H^\dagger i\overleftrightarrow{D_\nu}H)V^{\mu\nu}$ \\
\multicolumn{1}{c|}{} & \multicolumn{1}{c|}{} & $(\del_\mu\phi)(H^\dagger i\overleftrightarrow{D_\nu}H)\tilde V^{\mu\nu}$ \\
\multicolumn{1}{c|}{} & \multicolumn{1}{c|}{} & $(\del_\mu\phi)(H^\dagger i\overleftrightarrow{D_\nu}^A H)V^{A,\mu\nu}$ \\
\multicolumn{1}{c|}{} & \multicolumn{1}{c|}{} & $(\del_\mu\phi)(H^\dagger i\overleftrightarrow{D_\nu}^AH)\tilde V^{A,\mu\nu}$ \\ \hline
\multicolumn{1}{c|}{\multirow{2}{*}{$\phi H \bar H \psi_1 \bar \psi_2$}} & \multicolumn{1}{c|}{\multirow{2}{*}{$\asbraket{1}{p_\phi}{2}$}} & $(\del_\mu\phi)(\bar \psi\gamma^\mu\psi) |H|^2$ \\
\multicolumn{1}{c|}{} & \multicolumn{1}{c|}{} & $(\del_\mu\phi)(\bar \psi\gamma^\mu T^A\psi)( H^\dagger T^A H)$ \\ \hline
\multicolumn{1}{c|}{$\phi H_1\bar H_2 H_3 \bar H_4$} & \multicolumn{1}{c|}{$p_\phi\cdot (p_1-p_2)$ + symm.} & $(\del^\mu\phi)(H^\dagger i\overleftrightarrow{D_\mu}H)|H|^2$
\end{tabular}
\caption{Contact amplitudes suppressed by $1/f^3$ and corresponding $d=7$ operatores. The symbol $\psi$ denotes SM fermions, while $V=B$ and $V^A = W^A$, $G^A$ denote the abelian and non-abelian SM gauge bosons, respectively. Each operator must be invariant under the SM gauge group and this restricts the type of fields that can appear. The symbol `symm' indicates that it is necessary to symmetrize the momenta of identical particles according to Bose symmetry. Additional spinor structures can be obtained from the ones shown by swapping angle and square brackets. The operator $\protect\overleftrightarrow{D_\mu}^A$ is defined as $\protect\overleftrightarrow{D_\mu}^A=T^A \protect\overrightarrow{D_\mu} - \protect\overleftarrow{D_\mu}T^A$ and the dual field strength as $\tilde V_{\mu\nu} = \frac{1}{2}\epsilon_{\mu\nu\alpha\beta}V^{\alpha\beta}$.}\label{tab:dim7}
\end{table}

\begin{table}[t]\centering\renewcommand{\arraystretch}{1.5}
\begin{tabular}{ccc}
\multicolumn{3}{c}{Dimension 8} \\ \hline\hline
\multicolumn{1}{c|}{Particle content} & \multicolumn{1}{c|}{$\Amp^\text{ct} \times f^4$} & $\mathcal{O}\times f^4$ \\ \hline
\multicolumn{1}{c|}{$\phi^4$} & \multicolumn{1}{c|}{$(p_{\phi_1}\cdot p_{\phi_2})(p_{\phi_3}\cdot p_{\phi_4})$ + symm.} & $(\del_\mu\phi\,\del^\mu\phi)^2$ \\ \hline
\multicolumn{1}{c|}{\multirow{2}{*}{$\phi^2H_1\bar H_2$}} & \multicolumn{1}{c|}{$(p_{\phi_1}\cdot p_{\phi_2})^2$} & $(\del_\mu\del_\nu\phi\,\del^\mu\del^\nu\phi)|H|^2$ \\ \cline{2-3} 
\multicolumn{1}{c|}{} & \multicolumn{1}{c|}{$\left(p_{\phi_1}\cdot p_1\right)\left(p_{\phi_2}\cdot p_2\right)$ + symm.} & $(\del^\mu\phi\,\del^\nu\phi)(D_\mu H^\dagger D_\nu H)$ \\ \hline
\multicolumn{1}{c|}{\multirow{6}{*}{$\phi^2V_1V_2$}} & \multicolumn{1}{c|}{\multirow{4}{*}{$(p_{\phi_1}\cdot p_{\phi_2})\braket{12}^2$}} & $(\del_\alpha\phi\,\del^\alpha\phi)V_{\mu\nu}V^{\mu\nu}$ \\
\multicolumn{1}{c|}{} & \multicolumn{1}{c|}{} & $(\del_\alpha\phi\,\del^\alpha\phi)V_{\mu\nu}\tilde V^{\mu\nu}$ \\
\multicolumn{1}{c|}{} & \multicolumn{1}{c|}{} & $(\del_\alpha\phi\,\del^\alpha\phi)V^A_{\mu\nu}V^{A,\mu\nu}$ \\
\multicolumn{1}{c|}{} & \multicolumn{1}{c|}{} & $(\del_\alpha\phi\,\del^\alpha\phi)V^A_{\mu\nu}\tilde V^{A,\mu\nu}$ \\ \cline{2-3} 
\multicolumn{1}{c|}{} & \multicolumn{1}{c|}{\multirow{2}{*}{$\asbraket{1}{p_{\phi_1}}{2}\asbraket{1}{p_{\phi_2}}{2}$}} & $(\del^\mu\phi\,\del_\nu\phi)V_{\mu\alpha} V^{\alpha\nu}$ \\
\multicolumn{1}{c|}{} & \multicolumn{1}{c|}{} & $(\del^\mu\phi\,\del_\nu\phi)V^A_{\mu\alpha} V^{A,\alpha\nu}$ \\ \hline
\multicolumn{1}{c|}{$\phi^2\psi_1\bar \psi_2$} & \multicolumn{1}{c|}{$(p_{\phi_1}\cdot p_1)\asbraket{1}{p_{\phi_2}}{2}$ + symm.} & $(\del_\mu\phi\,\del_\nu\phi)(\bar \psi \gamma^\mu D^\nu \psi)$ \\ \hline
\multicolumn{1}{c|}{$\phi^2\psi_1\bar \psi_2 H$} & \multicolumn{1}{c|}{$(p_{\phi_1}\cdot p_{\phi_2})\braket{12}$} & $(\del_\mu\phi\, \del^\mu\phi)\bar \psi H \psi$ \\ \hline
\multicolumn{1}{c|}{$\phi^2 H^4$} & \multicolumn{1}{c|}{$p_{\phi_1}\cdot p_{\phi_2}$} & $(\del_\mu\phi\, \del^\mu\phi)|H|^4$
\end{tabular}
\caption{Same as Tables~\ref{tab:dim6} and~\ref{tab:dim7} for contact amplitudes of order $1/f^4$ and the correspondent $d=8$ operators. We do not consider amplitudes that violate baryon or lepton number.}\label{tab:dim8}
\end{table}

We will now systematically build amplitudes involving ALPs and SM particles in the unbroken electroweak phase, imposing Eq.\,\eqref{eq:Adler}. In practice, these amplitudes must scale with some positive power of the ALP momentum. This can either appear in the coefficients through Lorentz invariant combinations such as $p\cdot p_\phi$, or in the spinor structures themselves, for instance as $\asbraket{p}{p_\phi}{p'}$, with $p,p'$ the momenta of other particles in the amplitude. To have any chance to contribute to amplitudes suppressed at most by $f^4$, the spinor structures appearing must have dimension $\leq 4$. Additionally, we also have to impose Bose symmetry when there is more than one identical particle. For simplicity, we assume only one species of ALP $\phi$. 

Let us illustrate the method with an example, $\phi^2V_1V_2$, for which we have four helicity configurations for the vectors: $(-,-)$, $(+,+)$, $(+,-)$ and $(-,+)$. For the $(-,-)$ configuration, little-group covariance and dimensional analysis allow us to write only the spinor structure $\braket{12}^2$. This can satisfy Eq.\,\eqref{eq:Adler} only if multiplied by invariants proportional to $p_{\phi_{1,2}}$, appropriately symmetrized to satisfy Bose symmetry. The only possibility at the mass dimension of interest is $p_{\phi_1}\cdot p_{\phi_2}$ and the amplitude is
\be\label{eq:phi2V2_mm}
\Amp_4^\text{ct}\!\left[\phi^2 V_1^-V_2^-\right] \propto \frac{p_{\phi_1}\cdot p_{\phi_2}}{f^4}\braket{12}^2.
\ee
Note that we could have built different kinematical invariants with more powers of momenta, but they would induce an amplitude of higher order in $f$. An identical reasoning applies to the $(+,+)$ helicity configuration, for which the amplitude has exactly the same form, provided we exchange the angle brackets with square brackets. We thus obtain a second independent spinor structure. The $(-,+)$ configuration requires, instead, a momentum insertion of the ALP in between the brackets, as we need to connect an angle with a square spinor. The only symmetric combination in $p_{\phi_{1,2}}$ is then
\be\label{eq:phi2V2_mp}
\Amp_4^\text{ct}\!\left[\phi^2 V_1^-V_2^+\right] \propto \frac{1}{f^4}\asbraket{1}{p_{\phi_1}}{2}\asbraket{1}{p_{\phi_2}}{2}.
\ee
The $(+,-)$ configuration can be obtained from the equation above just swapping angle for squared brackets. Unlike the previous case, this does not produce an independent spinor structure since, due to Bose symmetry, once we exchange angle and square brackets, re-label the momenta as $1\leftrightarrow 2$ and use the identity $\sabraket{p}{q}{k}=\asbraket{k}{q}{p}$, we obtain exactly the same spinor combination we began with. As a consequence, for each vector $V$ we can reconstruct only 3 independent operators:
\al{
\La_{d=8}&\supset \frac{C}{f^4}(\del_\alpha\phi)(\del^\alpha\phi)V_{\mu\nu}V^{\mu\nu}+ \frac{C'}{f^4}(\del_\alpha\phi)(\del^\alpha\phi)V_{\mu\nu}\tilde V^{\mu\nu}+\frac{C''}{f^4}(\del^\mu\phi)(\del_\nu\phi)V_{\mu\alpha}  V^{\alpha \nu},
}
with $C,C',C''$ dimensionless coefficients. Note that in the list above we do not have the operator $(\del^\mu\phi)(\del_\nu\phi)V_{\mu\alpha} \tilde V^{\alpha \nu}$. Although not trivial, it can be shown that this operator is redundant\,\cite{Chala:2021cgt}. From our on-shell construction, however, the non-redundant operators were singled out automatically.\footnote{On-shell techniques have been used to systematically construct higher-dimensional operators in general effective theories\,\cite{Li:2022tec,DeAngelis:2022qco}.} We observe that, were we considering a generic pseudoscalar $\Phi$ and not an ALP $\phi$, it would be possible to write down additional terms in the amplitude, that vanish in the ALP case. For instance, for the case just discussed, the amplitude would be
\be
\Amp_4^\text{ct}[\Phi^2V_1^-V_2^-] = \frac{c_1}{f^2} \langle 12 \rangle^2 + \frac{c_1 \left(p_{\Phi_1}\cdot p_{\Phi_2}\right)}{f^4}\braket{12}^2 ,
\ee
where $c_{1,2}$ are dimensionless constants. It is only when we consider ALPs and we impose the ALP soft contact condition of Eq.\,\eqref{eq:Adler} that $c_1=0$ and we are left with Eq.\,\eqref{eq:phi2V2_mm}. As a second example, let us consider the amplitude between two pseudoscalars and two Higgs doublets. In the case of a generic pseudoscalar, the amplitude can be written up to $\order{f^{-4}}$ as
\be
\Amp_4^\text{ct}[\Phi^2 H_1 \bar H_2] = c_0' + \frac{c_1'}{f^2} p_{\Phi_1} \cdot p_{\Phi_2} + \frac{c_2'}{f^4} \left(p_{\Phi_1} \cdot p_{\Phi_2}\right)^2 + \frac{c_3'}{f^4} \left( (p_{\Phi_1} \cdot p_1 ) (p_{\Phi_2} \cdot p_2) + {\rm symm}\right),
\ee
where symm denotes the momenta symmetrization according to Bose symmetry and the constants $c_{0,1,2,3}'$ are dimensionless. On the other hand, when we consider ALPs, the ALP soft contact condition forces $c_0'=0$ and we are left only with the terms proportional to the ALPs momenta:
\be
\Amp_4^\text{ct}[\phi^2 H_1 \bar H_2] = \frac{c_1'}{f^2} p_{\phi_1} \cdot p_{\phi_2} + \frac{c_2'}{f^4} \left(p_{\phi_1} \cdot p_{\phi_2}\right)^2 + \frac{c_3'}{f^4} \left( (p_{\phi_1} \cdot p_1 ) (p_{\phi_2} \cdot p_2) + {\rm symm}\right).
\ee
For both cases of the generic pseudo-scalar and ALPs, the basis of amplitudes that satisfy the ALP soft contact condition \eqref{eq:Adler} is the same.

We present the list of higher-point amplitudes suppressed by $f^{-2}$, $f^{-3}$ and $f^{-4}$ and the correspondent dimension 6, 7 and 8 operators in Tables~\ref{tab:dim6}, \ref{tab:dim7} and~\ref{tab:dim8}, respectively. For each amplitude we only show independent spinor structures, with the exception of those that can be obtained exchanging angle with square brackets which are left implicit. We also do not write explicitly overall coefficients that may depend on SM gauge group structures like group generators.\footnote{In Eqs.\,\eqref{eq:phi2V2_mm} and \eqref{eq:phi2V2_mp}, for example, if $V$ is an abelian boson the proportionality factor is just a constant, while if it is non-abelian then $\Amp_4^\text{ct}[\phi^2 V^A_1V^B_2]\propto \delta^{AB}$, where $\delta^{AB}$ takes into account the $\mathcal{G}_\text{SM}$ tensor structure.} These coefficients will depend on the particles appearing in the amplitude and their quantum numbers. We show explicit examples in Appendix~\ref{app:conventions} for the case of 3-point amplitudes with SM particles. For the higher dimensional operators, we only show the general particle content, with $\psi$ representing SM fermions, $V=B$ the abelian SM gauge boson and $V^A = W^A$ or $G^A$ the SM non-abelian gauge bosons. Each operator should be invariant under the SM gauge group and, as usual, this imposes restrictions on the particles that can appear in each operators. We observe that some spinor structure involving spin 1 particles may involve both abelian or non-abelian gauge bosons. The difference in this case lies in the overall coefficient that we do not write explicitly, but the amplitude corresponds to different operators depending on the nature of the spin 1 particle involved. To avoid confusion, we make explicit the dependence on the gauge bosons when we write the operators corresponding to the amplitude under consideration. For instance, the $\langle 1 | p_{\phi_1} | 2 ] \langle 1 | p_{\phi_2} | 2]$ amplitude may correspond to the $d=8$ operators $(\partial^\mu \phi \partial_\nu \phi) V_{\mu\alpha} V^{\nu\alpha}$ or $(\partial^\mu \phi \partial_\nu \phi) V^A_{\mu\alpha} V^{A,\nu\alpha}$, depending on the vector considered.

We conclude stressing that the constant coefficients that multiply the amplitudes in Tables~\ref{tab:dim6}, \ref{tab:dim7} and~\ref{tab:dim8} are not further constrained by Eq.\,\eqref{eq:Adler}, in sharp contrast to the amplitudes suppressed by $f^{-1}$ discussed in Section~\ref{sec:SM}.

\subsection{Corrections to photon and gluon couplings}

The construction of amplitudes suppressed by higher powers of $f$ follows in similar fashion, with their number quickly growing. One interesting class of amplitudes is $\Amp^\text{ct}\!\left[\phi V V H^n\bar H^n\right]$, with $n\geq 1$ and $V$ a massless gauge boson, \textit{i.e.} a photon or a gluon at low-energy, that start only at dimension 7. In the non-dynamical limit of the Higgs, these amplitudes are the only ones that can modify the dimension 5 coupling of the 3-point amplitudes \eqref{eq:ALP_2vectors_SM_final} of the ALPs to massless gauge bosons, thus they can give us an insight of how the low-energy amplitudes are affected by the higher-point ones. 

According to little-group invariance, and taking the negative helicity configuration for the spin 1 particles as example, these amplitudes can be written schematically as
\be\label{eq:phiVV_HH}
\frac{\Amp^\text{ct}\!\left[\phi V_1^- V_2^- H^n\bar H^n\right]}{\braket{12}^2/f^{2n+1}} = c_0+(p_V\cdot p_H)\frac{c_1}{f^2}+(p_\phi\cdot p_H)\frac{c_2}{f^2}+(p_V\cdot p_\phi)\frac{c_3}{f^2}+\cdots,
\ee
where $c_{0,1,\cdots}$ are dimensionless (complex) constant coefficients and the dots denote higher-order terms with more powers of kinematical invariants. Moreover, $p_H$ represents Bose symmetric combinations of momenta of the vectors and Higgs, and $p_V=p_1+p_2$, which is the sole combination of $p_{1,2}$ that respects Bose symmetry. Imposing the ALP soft contact condition \eqref{eq:Adler} implies that $c_0=c_1=0$, while $c_2,c_3$ may be non-vanishing. As stressed before, we also need to take the non-dynamical limit of the Higgs\,\eqref{eq:higgsing} in order for $\Amp^\text{ct}\!\left[\phi V V H^n\bar H^n\right]$ to contribute to the low-energy amplitude $\Amp_3\!\left[\phi V V\right]$. Taking the non-dynamical limit\,\eqref{eq:higgsing} implies that all the structures in Eq.\,\eqref{eq:phiVV_HH} that contain $p_H$ will vanish at low-energy. Thus, the only relevant terms will be the ones proportional to $p_V\cdot p_\phi$. Since the resulting amplitude is a 3-point amplitude, momentum conservation implies that $p_V=-p_\phi$ and the final amplitude ends up proportional to $p_\phi^2$. All the steps above can be summarized as
\al{\label{eq:correction_phiVV}
\Amp^\text{ct}\!\left[\phi V_1^- V_2^- H^n\bar H^n\right] & \xrightarrow{\text{Eq.\,\eqref{eq:Adler}}}\left[(p_\phi\cdot p_H)\frac{c_2}{f^2}+(p_V\cdot p_\phi)\frac{c_3}{f^2}+\cdots\right]\frac{\braket{12}^2}{f^{2n+1}}\\
& \xrightarrow{\text{Eq.\,\eqref{eq:higgsing}}} \left[(p_V\cdot p_\phi)\frac{c_3}{f^2}+\cdots\right]\left(\frac{v}{f}\right)^{2n}\frac{\braket{12}^2}{f}\\
& \xrightarrow{p_V=-p_\phi} -\left[c_3+\cdots\right]\frac{p_\phi^2}{f^2}\left(\frac{v}{f}\right)^{2n}\frac{\braket{12}^2}{f},
}
where the term inside the square bracket is constant and can depend on higher powers of $p_\phi^2$. One the one hand, if the ALP is exactly massless, $p_\phi^2=0$, the low-energy couplings are not corrected by higher-dimensional operators, as Eq.\,\eqref{eq:correction_phiVV} vanishes. On the other hand, if we allow for a small mass $p_\phi^2=m_\phi^2$, we see that all corrections are suppressed by at least the ALP mass squared. This conclusion agrees with what was previously argued in Refs.\,\cite{Fraser:2019ojt,Agrawal:2017cmd,Reece:2023czb} and follows straightforwardly from our formalism.

\vspace{1em}
\section{A phenomenological application : $\ell^-\ell^+\to \phi h$}\label{sec:pheno}

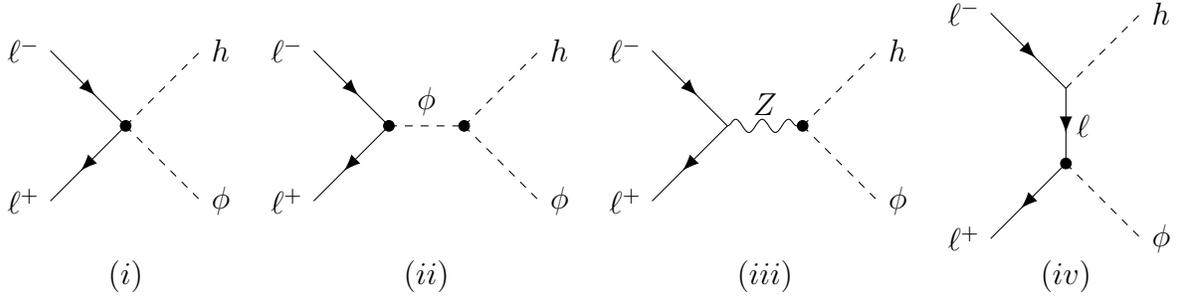
\begin{figure}
\begin{center}
    \begin{tikzpicture}
        \draw[f] (-1,1)node[left]{$\ell^-$} -- (0,0) ;
        \filldraw[black] (0,0) circle (2pt);
        \draw[f] (0,0) -- (-1,-1) node[left]{$\ell^+$};
        \draw[dashed] (0,0) -- (1,1) node[right]{$h$};
        \draw[dashed] (0,0) -- (1,-1) node[right]{$\phi$};
        \node[] at (0,-2) {$(i)$};
        \draw[f] (-1+3.5,1)node[left]{$\ell^-$} -- (0+3.5,0) ;
        \draw[f] (0+3.5,0) -- (-1+3.5,-1) node[left]{$\ell^+$};
        \draw[dashed] (0+3.5,0) -- (0+4.5,0) node[above,midway] {$\phi$};
        \filldraw[black] (0+3.5,0) circle (2pt);
        \filldraw[black] (0+4.5,0) circle (2pt);
        \draw[dashed] (0+4.5,0) -- (1+4.5,1) node[right]{$h$};
        \draw[dashed] (0+4.5,0) -- (1+4.5,-1) node[right]{$\phi$};
        \node[] at (4,-2) {$(ii)$};
         \draw[f] (7,1)node[left]{$\ell^-$} -- (8,0) ;
        \draw[f] (8,0) -- (7,-1) node[left]{$\ell^+$};
        \draw[snake it] (8,0) -- (9,0) node[above,midway] {$Z$};
        \filldraw[black] (9,0) circle (2pt);
        \draw[dashed] (9,0) -- (10,1) node[right]{$h$};
        \draw[dashed] (9,0) -- (10,-1) node[right]{$\phi$};
        \node[] at (8.5,-2) {$(iii)$};
        \draw[f] (11.5,1.5)node[left]{$\ell^-$} -- (12.5,0.5);
        \draw[dashed] (12.5,0.5) -- (13.5,1.5)node[right]{$h$};
        \draw[f] (12.5,0.5) -- (12.5, -0.5) node[right,midway] {$\ell$};
        \draw[f] (12.5,-0.5) -- (11.5, -1.5)node[left]{$\ell^+$};
        \filldraw[black] (12.5,-0.5) circle (2pt);
        \draw[dashed] (12.5,-0.5) -- (13.5,-1.5) node[right]{$\phi$};
        \node[] at (12.5,-2) {$(iv)$};
    \end{tikzpicture}
\end{center}
    \caption{Diagrams contributing to $\ell^-\ell^+ \to  \phi h$ generated by the operators described in the text. Black circles denote the insertion of operators containing the ALP. For the last diagram, also the crossed contribution in which the $h$ and $\phi$ legs are exchanged should be considered.}\label{fig:ll->phih}
\end{figure}

The phenomenological impact of higher-dimensional ALP operators have been previously considered in Refs.\,\cite{Bauer:2017ris,Bauer:2018uxu,Bauer:2016zfj,Bauer:2016ydr}, where the effects of the $d=6$ operator\linebreak $(\del_\mu\phi)(\del^\mu\phi)|H|^2$ and of the $d=7$ operator $(\del^\mu\phi)(H^\dag i \overleftrightarrow{D}_\mu H) |H|^2$ were studied in the context of collider physics. In this section we will turn our attention to the $d=7$ operators $(\partial_\mu \phi) (D^\mu \bar{\psi})H \psi$, $(\partial_\mu \phi) \bar{\psi}H (D^\mu \psi)$ and explore their phenomenology at lepton colliders. More precisely, we will study the impact of these operators in the process $\ell^- \ell^+ \to \phi h$, with $\ell=e,\mu$ and $h$ the physical Higgs, that can in principle be tested at future lepton colliders.

The full set of operators that we will consider is
\al{\label{eq:Lag_ll->hphi}
&\La_\text{int} = \frac{C_{\phi^2H^2}}{2f^2} (\del_\mu\phi)(\del^\mu\phi)|H|^2+\frac{C_{\phi H^4}}{f^3}(\del^\mu\phi)(H^\dag i \overleftrightarrow{D}_\mu H) |H|^2+\\
&+\sum_{\ell=e,\mu} i\frac{C_{\phi\ell^2 H}}{f}y_\ell\phi\bar L_\ell H \ell_R + \frac{C_{\phi\ell^2HD^2}^{(1)}}{f^3}(\del_\mu\phi)(D^\mu\bar L_\ell) H \ell_R + \frac{C_{\phi\ell^2HD^2}^{(2)}}{f^3}(\del_\mu\phi)\bar L_\ell H (D^\mu\ell_R) + h.c.,
}
where $L_{e,\mu}$ are the left-handed doublets of the first and second families, while $e_R$ and $\mu_R$ are the corresponding right-handed fields. We denote by $C_{\phi^2H^2},C_{\phi H^4},C_{\phi\ell^2H},C_{\phi\ell^2HD^2}^{(1,2)}$ the Wilson coefficients, noticing that $C_{\phi\ell^2H}$ is real and we have already factorized the Yukawa $y_\ell$ explicitly.\footnote{Having in mind the structure of Eq.\,\eqref{eq:coupling_SM_fermion}, our parametrization amounts to suppose that $Y_\ell \tilde{A}_R - \tilde{B}_R Y_\ell$
equals the Yukawa coupling times an order one factor that we denote by $C_{\phi\ell^2 H}$, as expected by power counting.} We do not consider lepton flavor violating couplings and will always assume that the $d=5$ coupling has the structure given by Eq.\,\eqref{eq:coupling_SM_fermion}. Also, we only consider ALP effective interactions and do not include any SMEFT operators. 

After electroweak symmetry breaking, the operators in Eq.\,\eqref{eq:Lag_ll->hphi} will generate the diagrams contributing to $\ell^-\ell^+\to \phi h$ shown in Fig.~\ref{fig:ll->phih}. More in detail, diagrams $(i)$ and $(iv)$ receive contributions from both $C_{\phi\ell^2H}$ and $C_{\phi\ell^2HD^2}^{(1,2)}$, diagram $(ii)$ gets contributions from $C_{\phi\ell^2H}$, $C_{\phi\ell^2HD^2}^{(1,2)}$ and $C_{\phi^2H^2}$ while diagram $(iii)$ is only generated by $C_{\phi H^4}$. To compute the corresponding cross-section, we generate and manipulate the total amplitude using {\tt FeynRules}\,\cite{Alloul:2013bka}, {\tt FeynArts}\,\cite{Hahn:2000kx} and {\tt FeynCalc}\,\cite{Shtabovenko:2016sxi,Shtabovenko:2020gxv}. In the limit of very high-energies, $\sqrt{s} \gg v$, with $\sqrt{s}$ the center-of-mass energy and $v$ the Higgs vacuum expectation value, the differential cross-section in the center-of-mass frame simplifies to the following expression:
\begin{figure}[t]
\centering
\includegraphics[width=0.47\textwidth]{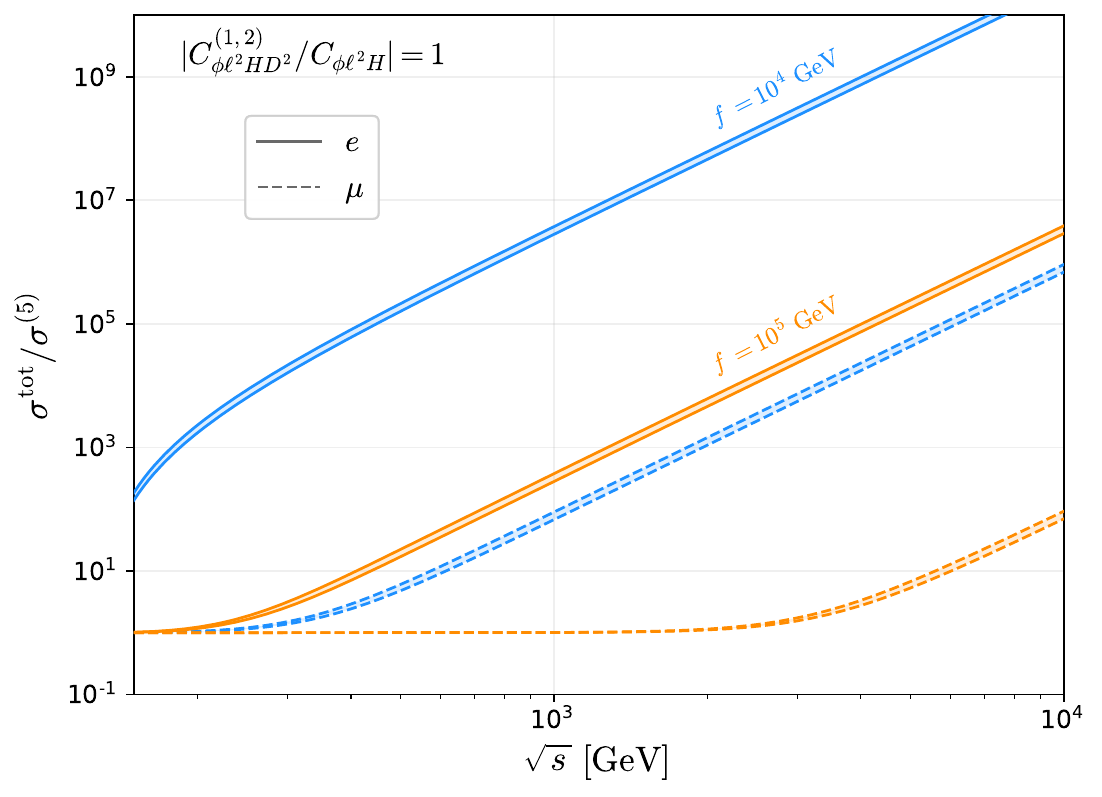}
\includegraphics[width=0.47\textwidth]{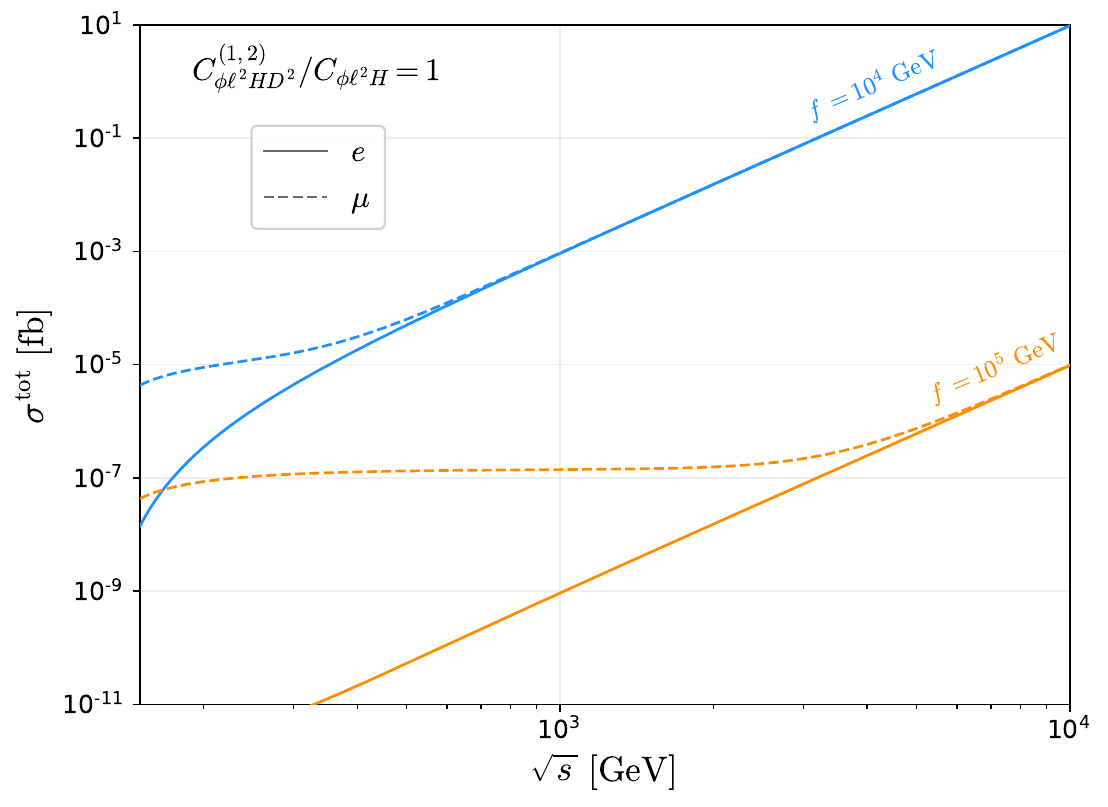}
\caption{\textit{Left}: Ratio between the cross-sections with and without dimension 7 operators as a function of the center of mass energy. We take the modulus of the ratio of $d=7$ and $d=5$ coefficients equal to $1$. The bands are obtained by varying the sign between $C_{\phi\ell^2HD^2}^{(1,2)}$. \textit{Right}: Total cross-section with the inclusion of $d=7$ operators as a function of the center of mass energy, taking $C_{\phi\ell^2HD^2}^{(1,2)}=C_{\phi\ell^2H}=1$. All cross-sections are computed taking $m_\phi=0$. For both panels the solid curves denote $\ell=e$ and dashed $\ell=\mu$, while blue (orange) means $f=10^4$ ($10^5$) GeV.}
\label{fig:xsec}
\end{figure}
\al{\label{eq:ll_hphi_xsec}
\frac{\dd \sigma^\text{tot}(\ell^-\ell^+\to \phi h)}{\dd \cos\theta}&\simeq  \frac{1}{512\pi f^6}\Big[(u^2+t^2)\left(|C_{\phi\ell^2HD^2}^{(1)}|^2+|C_{\phi\ell^2HD^2}^{(2)}|^2\right)+\\
&\qquad\qquad + 4tu\text{Re}\left(C_{\phi\ell^2HD^2}^{(1)}C_{\phi\ell^2HD^2}^{(2)*}\right)\Big],
}
where $t=(p_{\ell^-}-p_{h})^2$ and $u=(p_{\ell^-}-p_\phi)^2$. Inspection of Eq.\,\eqref{eq:ll_hphi_xsec} shows that, in the high energy limit, the only relevant contributions are those coming from $C_{\phi\ell^2HD^2}^{(1,2)}$ and, more specifically, from diagram $(i)$, due to the fact that only this diagram give a contribution that grows quadratically with the center-of-mass energy. The conclusion is that, even though $d=7$ operators are suppressed by more powers of $f$, they can still give the dominant contribution to observables for sufficiently high energies. This strongly relies on the fact that the coefficient of the $d=5$ operator is Yukawa suppressed, while an analogue suppression does not exist for the $d=7$ operators. We also observe that Eq.\,\eqref{eq:ll_hphi_xsec} is still the leading contribution even if effects from $d=8,9$ operators are included. This is because, to give a cross section at the same order in $1/f$, they would need to interfere with $d=6,5$ operators, respectively, and would thus be suppressed by powers of masses and $v$. The same is true if loop effects involving multiple insertions of the $d=5$ operator are considered, since they are doubly suppressed (by one loop factor and by the lepton Yukawa coupling) and are thus not expected to change our conclusions.

To quantify the effects of the $d=7$ operators and compare them to those of the $d=5$ operator, we show in Fig.~\ref{fig:xsec} the complete cross-section (in which particle masses and the Higgs vacuum expectation value are properly taken into account) as a function of $\sqrt{s}$, with $\sigma^{\text{tot},(5)}$ denoting, respectively, the total cross-section and the cross-section computed with only $d=5$ operators.
Since $C_{\phi\ell^2HD^2}^{(1,2)}$ give the dominant contribution at high-energies, we set $C_{\phi^2H^2}=C_{\phi H^4}=0$. Furthermore, we take $C_{\phi\ell^2H}=|C_{\phi\ell^2HD^2}^{(1,2)}|=1$, while we allow for a different signs between $C_{\phi\ell^2HD^2}^{(1)}$ and $C_{\phi\ell^2HD^2}^{(2)}$. On the left panel, we plot the ratio $\sigma^{\text{tot}}/\sigma^{(5)}$ for $\ell=e,\mu$ and $f=10^4$ GeV (blue lines) and $f=10^5$ GeV (orange lines). Continuous lines refer to $e^+ e^- \to  \phi h$ while dashed lines refer to $\mu^+ \mu^- \to  \phi h$. The bands are obtained varying the sign between $C_{\phi\ell^2HD^2}^{(1)}$ and $C_{\phi\ell^2HD^2}^{(2)}$. As we can see, at high energies $\sigma^\text{tot}$ can be larger than $\sigma^{(5)}$ by many orders of magnitude. The effect is larger for electrons, since $\sigma^{(5)}$ is suppressed by smaller Yukawa couplings. 

On the right panel we instead show how $\sigma^\text{tot}$ grows with energy, with the same conventions used in the left panel. Inspecting the two muon cross-sections shown, we clearly see the transition from the low energy regime (dominated by the $d=5$ operator and essentially independent on energy \footnote{When $\sqrt{s}$ is comparable to the Higgs mass, the cross section is not flat with respect to $\sqrt{s}$ because of the threshold.}) and the high energy energy regime, in which the $d=7$ operator dominates and the cross-section scales as $s^2$ as predicted by\linebreak Eq.\,\eqref{eq:ll_hphi_xsec}. It is interesting to notice that for $f=10^4$ GeV the dimension 7 effects can push the cross-section to reasonable values ($\sigma^\text{tot}\sim 1$ fb), while it would remain out of reach considering only the $d=5$ operator ($\sigma^{(5)}\lesssim 10^{-5}$ fb).

From the EFT perspective, it is to be expected that higher-dimensional operators dominate the cross-section when the energy is close to the cut-off $f$. What makes the present case more special and interesting is that the effects of the $d=7$ operators are relevant much before $\sqrt{s}\sim f$. This is a direct consequence of the fact that the ALP soft factorization condition \eqref{eq:soft_factor} required the coefficient of the $d=5$ amplitude to be proportional to the Yukawa, while the ALP soft contact condition \eqref{eq:Adler} did not impose such selection rule on the $d=7$ operators. We see from Fig.~\ref{fig:xsec} that, at the phenomenological level, this distinction between $d=5$ and $d=7$ operators might be extremely relevant, since, for instance for electrons, the contribution from $d=5$ is always subdominant.

\section{Conclusions}\label{sec:conclusion}

Given a scattering amplitude involving ALPs, what is the set of physical properties that it must satisfy to recover shift-symmetry? Of course, starting from a Lagrangian and requiring it to be shift-symmetry invariant is not a complete answer to this question, as we are not making statements about the amplitudes themselves. Taking the on-shell approach, that bypasses fields and Lagrangians, makes the question even less obvious, but provides a natural framework for working it out. In this paper, using on-shell methods, we have identified three conditions that allow us to construct amplitudes with the desired shift-symmetric properties. All three conditions rely on the properties of the amplitude in the limit of soft ALP momentum.


The first condition, that we call \textit{ALP soft factorization condition}, Eq.\,\eqref{eq:soft_factor}, enables to reproduce the correct amplitudes when the ALP is interacting with two massive particles, {\it i.e.}, in the case of amplitudes suppressed by $f^{-1}$, with $f$ the scale associated with the ALP (Section~\ref{sec:toy}). In addition to constructing these amplitudes, we also discuss at length how they can be connected with amplitudes that involve SM particles and are invariant under the SM symmetry group at high energies. While the case of the coupling of an ALP to Higgs doublets or gauge bosons is relatively straightforward, the case of fermions is more subtle, due to an interesting obstruction that emerges because of a combination of the kinematic of 3-point amplitudes and invariance under the SM symmetry group. In Section~\ref{sec:SM} we identify a second condition, the \textit{ALP-Higgs obstruction}, that deals with this latter amplitude by relying on electroweak symmetry breaking and on the fact that, in the limit in which the Higgs momentum becomes soft, the 4-point amplitude which is invariant under the SM symmetry becomes a 3-point amplitude with massive fermions.
The consequence is that the scattering amplitude ends up being proportional to the Yukawa coupling of the fermions involved.

The third condition, that we call {\it ALP soft contact condition}, Eq.\,\eqref{eq:Adler}, allows to determine the correct amplitudes when the ALP is interacting with more than two other particles. Since these higher-point amplitudes correspond to effective operators suppressed by higher powers of $f$, we determine in Section~\ref{sec:higher}, solely using scattering amplitudes, the complete list of operators up to $d=8$. Among the higher dimensional operators, of particular interest for low energy physics are those that involve one ALP, two massless spin-1 bosons and a number of Higgs pairs, since in the limit of non-dynamical Higgs they correct the dimension 5 amplitude between one ALP and two massless spin-1 bosons. We show that on-shell techniques allow us to conclude in a straightforward way that the dimension 5 amplitude is corrected only if the ALP mass is different from zero, while the corrections vanish for a massless ALP.
We also study a phenomenological application in which these higher dimensional operators can dominate over the $d=5$ ones: the process $\ell^+ \ell^- \to  \phi h$ at high energy lepton colliders (Section~\ref{sec:pheno}). We find that, due to the Yukawa suppression of the $d=5$ operator, the $d=7$ ones can dominate already at energies $\sqrt{s} \ll f$, where they are typically expected to give only subdominant contributions with respect to the $d=5$ operators.

We observe that the \textit{ALP soft factorization condition} and the \textit{ALP soft contact condition} are equivalent to requiring the amplitude to manifest the Adler's zero in the limit of soft ALP limit. Nevertheless, this procedure is not completely universal, since they cannot be applied to 4-point amplitudes involving one Higgs doublet. As we have shown, in this case we must resort to the \textit{ALP-Higgs obstruction} condition of\linebreak Eq.\,\eqref{eq:Higgs_obstruction}, which can be seen as a generalization of the Adler's zero to this case.

Our work can be extended in several directions. First of all, having derived the shift-symmetric ALP amplitudes, it follows immediately that all additional terms that do not satisfy our conditions must break the shift-symmetry. An interesting point would be to investigate how some sort of power counting could be applied to such coefficients without resorting to Lagrangians. Moreover, as we already conjectured at the end of Section\,\ref{sec:toy}, it would be interesting to study how general loop diagrams could make regular the multi-soft limit in which several ALPs momenta are taken to zero at the same time. 
A third aspect that can be explored would be the generalization to amplitudes involving several different ALPs, {\it i.e.} to the case in which we have spontaneous symmetry breaking of a non-abelian group. Possible applications would be to the study of amplitudes involving light mesons and matter particles (like nucleons or vector mesons) in Chiral Perturbation Theory, or the coupling between a Composite Higgs and SM fermions and gauge bosons. Moreover, since on-shell methods can be used to derive the renormalization group equations of effective operators\,\cite{Baratella:2020lzz,EliasMiro:2020tdv,Baratella:2020dvw,Jiang:2020mhe}, amplitudes with more than one ALP can be used to study the effects of light ALPs in the running of Wilson coefficients of operators of the Standard Model Effective Field Theory. Finally, the on-shell scattering amplitudes formalism extended using the techniques of Ref.\,\cite{Csaki:2020inw} can, in principle, be used to compute the interactions between ALPs and photons when magnetic monolopoles are present\,\cite{Sokolov:2022fvs,Sokolov:2023pos}. We leave the study of these aspects to future work.

\vspace{1cm}


\acknowledgments

We thank M.~Ramos, G.~Guedes, J.~Roosmale Nepveu, and specially C.-Y.~Yao and J.~Kley for useful discussion. We are also grateful to S.~De~Angelis, Q.~Bonnefoy, G.~Durieux and Y.~Shadmi for insightful  comments on an early draft.
E.B. acknowledges financial support from "Fundação de Amparo à Pesquisa do Estado de São Paulo" (FAPESP) under contract 2019/04837-9, as well as partial support from ``Conselho Nacional de Deselvolvimento Cient\'ifico e Tecnol\'ogico'' (CNPq).
G.M.S. acknowledges financial support from "Fundação de Amparo à Pesquisa do Estado de São Paulo" (FAPESP) under contracts 2020/14713-2 and 2022/07360-1.
This work is supported by the Deutsche Forschungsgemeinschaft under Germany’s Excellence Strategy EXC 2121 “Quantum Universe” - 390833306. This work has been also partially funded by the Deutsche Forschungsgemeinschaft under the grant 491245950. This project has received funding from the European Union’s Horizon Europe research and innovation programme under the Marie Skłodowska-Curie Staff Exchange grant agreement No 101086085 - ASYMMETRY.
This project has received funding /support from the European Union’s Horizon 2020 research and innovation programme under the Marie Skłodowska -Curie grant agreement No 860881-HIDDeN.

\appendix
\section{Conventions}\label{app:conventions}

\subsection{Spinor variables}

In this Appendix, we set the notation for the spinor variables and summarize the identities we use. The starting point is the covariance of the $S$-matrix under little-group, that allows us to write scattering amplitudes as a sum of all possible kinematical structures that carry the correct little-group transformation, each multiplied by a corresponding coupling. The building blocks for these kinematical structures are 2-component spinors. For massless momenta we define the spinor variables as
\be
\ket{p}\equiv \ket{p}_\alpha,\quad \sket{p}\equiv \sket{p}^{\dalpha},\quad \bra{p}\equiv \bra{p}^\alpha,\quad \sbra{p}\equiv \sbra{p}_{\dalpha},
\ee
where $\alpha,\dalpha$ are $SL(2,C)$ indices for left- and right-handed spinors, respectively. The spinors $\ket{p},\bra{p}$ are referred to as \textit{angle} spinors, while $\sket{p},\sbra{p}$ as \textit{square} spinors. They transform under the $U(1)$ little-group with opposite phases:
\be\label{eq:LG_massless}
\ket{p}\to e^{-i\eta}\ket{p},\quad \sket{p}\to e^{i\eta}\sket{p},\quad \eta\in\mathbb{R}.
\ee
All indices can be raised and lowered with the Levi--Civita tensor that is defined by $\epsilon^{12}=-\epsilon_{12}=1$, for instance $\ket{p}_\alpha = \epsilon_{\alpha\beta}\bra{p}^\beta$. The spinors above satisfy
\be\label{eq:momentum_spinors_massless}
p_\mu (\bar \sigma^\mu)^{\dalpha \alpha}\equiv p^{\dalpha \alpha} \equiv  \sket{p}^{\dalpha}\bra{p}^\alpha,\quad p_\mu (\sigma^\mu)_{\alpha \dalpha}\equiv p_{\alpha \dalpha} \equiv \ket{p}_{\alpha}\sbra{p}_{\dalpha},
\ee
\be\label{eq:Weyl_massless}
p\ket{p}=p\sket{p}=\bra{p}p=\sbra{p}p=0.
\ee
Equation \eqref{eq:momentum_spinors_massless} is simply the defining equation of the spinors, where $\sigma^\mu=(\bm{1},\vec{\sigma})$ and $\bar\sigma^\mu=(\bm{1},-\vec{\sigma})$, while in Eq.\,\eqref{eq:Weyl_massless} we have the Weyl equations. For a massless 4-momentum $(p^\mu)=|\vec p\,|(1,\sin\theta\cos\phi,\sin\theta\sin\phi,\cos\theta)$, with $\vec p$ the 3-momentum, one possible realization for the spinors in Eq.\,\eqref{eq:momentum_spinors_massless} is given by
\al{\label{eq:SpinorVariables_massless}
(\bra{p}^\alpha) = -\sqrt{2|\vec p\,|}\begin{pmatrix}  c \\  s^* \end{pmatrix},&\quad  (\sket{p}^{\dot \alpha}) = -\sqrt{2|\vec p\,|}\begin{pmatrix}  c \\  s \end{pmatrix},\\
(\ket{p}_\alpha) = \sqrt{2|\vec p\,|}\begin{pmatrix}  s^* \\  -c \end{pmatrix},&\quad   (\sbra{p}_{\dot \alpha}) = \sqrt{2|\vec p\,|}\begin{pmatrix}  s \\  -c \end{pmatrix},}
where $c\equiv \cos\frac{\theta}{2}$, $s\equiv e^{i\phi}\sin\frac{\theta}{2}$. We can also define the usual anti-symmetric Lorentz invariant products of spinors:
\be
\braket{pq}\equiv\bra{p}^\alpha\ket{q}_\alpha,\quad  \sbraket{pq}\equiv\sbra{p}_{\dalpha}\sket{q}^{\dalpha},
\ee 
with the contraction of up- and down-indices performed with the Levi--Civita tensor. These products can be computed explicitly using Eq.\,\eqref{eq:SpinorVariables_massless}.

For massive momenta we adopt the notation of Ref.\,\cite{Arkani-Hamed:2017jhn} and define bold angle and square spinors as
\al{
\ket{\bm{p}}\equiv \ket{p^I}_\alpha,\quad \sket{\bm{p}}\equiv \sket{p^I}^{\dalpha},\quad \bra{\bm{p}}\equiv \bra{p^I}^\alpha,\quad \sbra{\bm{p}}\equiv \sbra{p^I} _{\dalpha},
}
where now $I=1,2$ is the $SU(2)$ index of the little-group, that are also raised and lowered through the Levi--Civita tensor. Contrary to the massless case, the bold spinors transform in the same way under little-group as
\be\label{eq:LG_massive}
\ket{p^I}\to W^I_{\ J}\ket{p^J},\quad \sket{p^I}\to W^I_{\ J}\sket{p^J},\quad W\in SU(2),
\ee
and analogous for $\bra{\bm{p}}$ and $\sbra{\bm{p}}$. The massive spinors satisfy similar relations as\linebreak Eq.\,\eqref{eq:momentum_spinors_massless}, but with the inclusion of the $SU(2)$ little-group indices:
\be\label{eq:momentum_spinors}
p^{\dalpha \alpha} \equiv  \epsilon_{IJ}\sket{p^I}^{\dalpha}\bra{p^J}^\alpha,\quad  p_{\alpha \dalpha} \equiv  -\epsilon_{IJ}\ket{p^I}_{\alpha}\sbra{p^J}_{\dalpha}.
\ee
With the same parametrization as in Eq.\,\eqref{eq:SpinorVariables_massless}, we can represent the massive spinors in terms of the components of the 4-momentum as
\al{\label{eq:SpinorVariables_massive}
(\bra{p^I}^\alpha) = -\left(\begin{array}{c|c}
c\sqrt{E+|\vec p\,|} & - s\sqrt{E-|\vec p\,|}\\
s^*\sqrt{E+|\vec p\,|} & c\sqrt{E-|\vec p\,|}
\end{array}\right),&~ 
(\sket{p^I}^{\dot \alpha}) = -\left(\begin{array}{c|c}
s^*\sqrt{E-|\vec p\,|} & c\sqrt{E+|\vec p\,|}\\
-c\sqrt{E-|\vec p\,|} & s\sqrt{E+|\vec p\,|}
\end{array}\right),\\
(\ket{p^I}_\alpha) = \left(\begin{array}{c|c}
s^*\sqrt{E+|\vec p\,|} &  c\sqrt{E-|\vec p\,|}\\
-c\sqrt{E+|\vec p\,|} & s\sqrt{E-|\vec p\,|}
\end{array}\right),&~ 
(\sbra{p^I}_{\dot \alpha}) = \left(\begin{array}{c|c}
-c\sqrt{E-|\vec p\,|} & s\sqrt{E+|\vec p\,|}\\
-s^*\sqrt{E-|\vec p\,|} &- c\sqrt{E+|\vec p\,|}
\end{array}\right),
}
where the first (second) column refers to $I=1$ (2), while the rows are for different $SL(2,C)$ indices. Also, $E=\sqrt{m^2+|\vec p\,|^2}$ is the energy and, as before, $\vec p$ the 3-momentum. In writing Eq.\,\eqref{eq:SpinorVariables_massive}, we have assumed that momenta, and in particular the mass $m$, are real and positive. For the expressions with complex momenta, we refer the reader to Ref.\,\cite{Arkani-Hamed:2017jhn} and references therein. From the expressions above it is immediate to check that
\al{\label{eq:Weyl}
p_{\alpha \dot \alpha} \sket{p^I}^{\dot \alpha} = - m \ket{p^I}_\alpha, &\quad p^{\dot \alpha  \alpha} \ket{p^I}_{\alpha} = - m \sket{p^I}^\alpha,\\
\bra{p^I}^\alpha p_{\alpha \dot \alpha} = m \sbra{p^I}_{\dot \alpha},&\quad \sbra{p^I}_{\dot \alpha} p^{\dot \alpha \alpha} = m \bra{p^I}^{ \alpha},
}
which are the massive version of the Weyl equations in \eqref{eq:Weyl_massless}. We note that in the massive Weyl equations \eqref{eq:Weyl} the two types of spinors are related by the mass $m$. The anti-symmetric spinor product is defined in the same way as before
\be\label{eq:spinor_products}
\braket{\bm{pq}}\equiv \braket{p^Iq^J}=\bra{p^I}^\alpha\ket{q^J}_\alpha,\quad  \sbraket{\bm{pq}}\equiv \sbraket{p^Iq^J}=\sbra{p^I}_{\dalpha}\sket{q^J}^{\dalpha}.
\ee 
From the Weyl equations we can also derive
\be\label{eq:equal_braket}
\braket{p^Ip^J}=-m\epsilon^{IJ},\quad \sbraket{p^Ip^J}=m\epsilon^{IJ}.
\ee

It is possible to connect the spinor variables associated to massive particles to the corresponding massless variables. Each of the massive spinor helicity variables in Eq.\,\eqref{eq:SpinorVariables_massive} can be decomposed as\,\cite{Arkani-Hamed:2017jhn}
\be
\ket{p^I} = \ket{p} \zeta^I_+ + \sqrt{m} \sket{\hat p} \zeta^I_-, ~~~ \sket{p^I} = \sket{p} \zeta_+^I + \sqrt{m} \ket{\hat p} \zeta_-^I,
\ee
where $\ket{p},\sket{p}$ are the same massless variables in Eq.\,\eqref{eq:SpinorVariables_massless} and $\zeta_+ = (1,0)$, $\zeta_- = (0,1)$. This decomposition is the mathematical expression of our physical intuition that in the massive spinor helicity variables there will always be a component with the ``wrong'' helicity (\textit{i.e.} angle brackets will contain square brackets and vice-versa), but this contribution will be proportional to the particle mass. To make explicit the mass dependence, we have introduced the dimensionless spinor helicity variables $\ket{\hat p} \equiv \ket{p}/\sqrt{m}$ and $\sket{\hat p} \equiv \sket{p}/\sqrt{m}$. To see why this decomposition is useful, we discuss the example of the product $\langle \bm{pq} \rangle = \langle p^I q^J \rangle$, that can be written as
\be
\langle \bm{pq} \rangle = \langle pq \rangle \left(\zeta_+^I \zeta_+^J + \frac{m}{\langle pq \rangle} [ \hat{p} \hat{q}] \zeta_-^I \zeta_-^J \right).
\ee
Given that the product $\sbraket{\hat p \hat q}$ is finite, we define the \textit{high-energy limit} as the limit $m/\braket{pq} \to 0$, for which the second term between brackets vanishes. In a more concise way, the high-energy limit of $\braket{\bm{pq}}$ is given by
\be
\lim_{\rm H.E.} \braket{\bm{pq}} =\lim_{\rm H.E.} \braket{ p^I q^J } = \braket{pq} \delta^I_1 \delta^J_1\, ,
\ee
where $\rm H.E.$ stands for High-Energy, \textit{i.e.} for the $m/\braket{ pq} \to 0$ limit, and $\delta_1^I \delta_2^J = \zeta_+^I \zeta_+^J$. An analogous discussion also holds for $\sbraket{\bm{pq}}$. We thus see that, for the amplitudes with a simple dependence on the spinor helicity variables as those we are discussing in this paper, the high energy limit simply amounts to the ``unbolding'' of the massive spinor products
\be\label{eq:unbolding}
\braket{\bm{pq}}\to \braket{pq}\delta^I_1\delta^J_1,\quad \sbraket{\bm{pq}}\to \sbraket{pq}\delta^I_2\delta^J_2.
\ee

It is also useful to consider spinors with negative momenta, for which the analytic continuation compatible with Eq.\,\eqref{eq:Weyl} reads
\be\label{eq:analytic_cont}
\ket{-\bm{p}}=\ket{\bm{p}},\quad \sket{-\bm{p}}=-\sket{\bm{p}}.
\ee 
Other useful identities are
\al{
\asbraket{p}{q}{k} = &\sabraket{k}{q}{p} ,\\
 \asbraket{p}{\sigma^\mu}{q}\asbraket{k}{\sigma_\mu}{l} &= -2\braket{pk}\sbraket{ql}
}
where $\asbraket{p}{q}{k}\equiv \bra{p}^\alpha q_{\alpha\dalpha} \sket{k}^{\dalpha}$ and similar for $\sabraket{k}{q}{p}$. The same relations also hold for massive spinors.

Finally, to compute the soft factors in Section \ref{sec:toy}, more precisely the $p_\phi$-independent piece in Eq.\,\eqref{eq:S_psi}, we can use the explicit representations for the spinors in Eqs.\,\eqref{eq:SpinorVariables_massless} and \eqref{eq:SpinorVariables_massive}. Starting with a massive momentum $p$ and a massless one $p_\phi$, we have:
\al{\label{eq:collinear_product_massive}
\braket{p^I~(p+p_\phi)^J} &= \begin{pmatrix}
0 & -m \\
\sqrt{(E-|\vec p\,|)(E+|\vec p\,|+2|\vec p_\phi|)} & 0
\end{pmatrix}^{IJ},\\
\sbraket{p^I~(p+p_\phi)^J} &= \begin{pmatrix}
0 & \sqrt{(E-|\vec p\,|)(E+|\vec p\,|+2|\vec p_\phi|)} \\
-m & 0
\end{pmatrix}^{IJ},
}
where we are assuming that the angle between the 3-momenta $\vec p$ and $\vec p_\phi$ is zero. We can see that in the limit $p_\phi\to0$ we recover Eq.\,\eqref{eq:equal_braket}. Had we instead started with both massless $p$ and $p_\phi$, such that the angle between the 3-momenta is $\theta\ll 1$, we would have obtained
\al{\label{eq:collinear_product_massless}
\braket{p~(p+p_\phi)^I} &=\theta\left(-|\vec p_\phi|\sqrt{\frac{|\vec p\,|}{|\vec p\,|+|\vec p_\phi|}},~-|\vec p\,|\sqrt{\frac{|\vec p_\phi|}{|\vec p\,|+|\vec p_\phi|}}~\right)^I+\order{\theta^2},\\
\sbraket{p~(p+p_\phi)^I} &=\theta\left(-|\vec p\,|\sqrt{\frac{|\vec p_\phi|}{|\vec p\,|+|\vec p_\phi|}},~|\vec p_\phi|\sqrt{\frac{|\vec p\,|}{|\vec p\,|+|\vec p_\phi|}}~\right)^I+\order{\theta^2},
}
which are relevant quantities for the computation of Eq.\,\eqref{eq:S_psi_many}. Note that in Eq.\,\eqref{eq:collinear_product_massless} we still have a $SU(2)$ little-group index, as $p+p_\phi$ is not massless.

\subsection{Construction of amplitudes}

To write down amplitudes, we use the fact that the $S$-matrix is covariant under little-group transformations. This implies that we can express the amplitudes as a sum of all possible Lorentz-invariant combinations of spinor variables that have the correct little-group transformation.

For massless particles the little-group transformation is given by Eq.\,\eqref{eq:LG_massless} and, according to the covariance of the $S$-matrix, the amplitude must transform as $\Amp\to e^{-2ih}\Amp$, with $h$ the helicity of the corresponding particle. Specializing to the case of 3-point amplitudes, little-group covariance gives us three constraints, while the amplitude can be built out of 6 spinor products, namely $\braket{12},\braket{23},\braket{13}$ and the same with square brackets. We use here the short-hand notation $\ket{n}\equiv\ket{p_n}$ to label the momenta in the spinors. Due to 3-particle kinematics, however, all these products vanish if the momenta are real. Relaxing this hypothesis, we find that we can either have the angle contractions or the square ones to be non-vanishing, thus leaving us with only three independent Lorentz-invariant products. Up to coefficients, this allows us to completely fix the amplitude\,\cite{Henn:2014yza,Elvang:2013cua,Dixon:2013uaa}:
\al{
\Amp_3 \propto\left\{\begin{array}{ll}
\braket{12}^{h_3-h_1-h_2}\braket{23}^{h_1-h_2-h_3}\braket{13}^{h_2-h_1-h_3},\quad& \sum_i h_i\leq 0\\
\sbraket{12}^{h_1+h_2-h_3}\sbraket{23}^{h_2+h_3-h_1}\sbraket{13}^{h_1+h_3-h_2},\quad& \sum_i h_i\geq0
\end{array}\right.,
}
where $h_{1,2,3}$ are the helicities of the particles. For higher-point functions, we cannot completely fix the amplitude, since we have more spinor contractions than little-group transformation rules.

When considering massive particles the discussion is more involved, because both angle and square spinors transform in the same way under little-group. Nevertheless, one can simply write the amplitude as a sum of all independent spinor structures, each with a different coefficient (see Ref.\,\cite{Durieux:2020gip} for a classification). For a massive particle of spin $s$, the transformation of the amplitude is given by the completely symmetric $2s$ tensor representation, which is equivalent to the usual representation in terms of the total spin and its projection\,\cite{Arkani-Hamed:2017jhn}. Hence, each term in the amplitude must contain exactly $2s$ spinors, angle and/or square, of this given particle.




\subsection{\textit{CPT} invariance and unitarity}

To extract information on the phases of the amplitudes, we need to relate them to their complex conjugate, which can be done using $CPT$ invariance and unitarity. In terms of the transfer matrix $T$, the amplitude for a state $\mathcal{O}$ is written as $\Amp\left[\mathcal{O}\right] = \braket{0|T\mathcal{O}}$, where we take all particles to be incoming and $0$ is the vacuum. Using $CPT$ invariance of the $S$-matrix, one can show that $\braket{0|T\mathcal{O}}=\braket{\mathcal{O}_\Theta|T0}$, with $\Theta$ representing the action of $CPT$ in the multi-particle states, that amounts to reversing the spin (\textit{i.e.} changing up and down little-group indices in the massive case, or flipping the helicity for massless particles) and swapping particles with anti-particles\,\cite{Weinberg:1995mt}. In addition, when applying $\Theta$ we must also reverse the ordering of the particles in the amplitude, that can lead to extra minus signs in the case of fermions. Then, unitarity of the $S$-matrix imposes that $T\simeq T^\dagger$ up to corrections of order $T^\dagger T$. We therefore obtain at leading order
\be\label{eq:CPT_amp}
\Amp\left[\mathcal{O}\right]\simeq \Amp\left[\mathcal{O}_\Theta\right]^*.
\ee
The complex conjugation of spinor structures can be found to yield
\be
\braket{p^Iq^J}^*=-\sbraket{p_Iq_J},
\ee
that holds when considering real momenta with positive energy\,\cite{Gunion:1985vca,Dixon:1996wi}.

\subsection{Feynman rules with spinor variables}

It is also instructive to show how some Feynman rules for effective operators are translated in terms of spinor variables. To do so, we first need to express the external wave-functions in terms of spinors. For spin 1/2 and spin 1 particles we have, respectively,
\al{
v^I(p) = \begin{pmatrix} \ket{p^I} \\ \sket{p^I} \end{pmatrix},&\quad \bar v_I(p) = \left( -\bra{p_I},~\sbra{p_I} \right),\\
u^I(p) = \begin{pmatrix} \ket{p^I} \\ -\sket{p^I} \end{pmatrix},&\quad \bar u_I(p) = \left( \bra{p_I},~\sbra{p_I} \right),
}
\be
\epsilon_\mu(p) = \frac{\asbraket{\bm{p}}{\sigma_\mu}{\bm{p}}}{\sqrt{2} m_V},\quad\text{or}\quad \epsilon(p)_{\alpha\dot{\alpha}} = \sqrt{2}\frac{\ket{\bm{p}}\sbra{\bm{p}}}{m_V},
\ee
where $u,v,\bar u,\bar v$ are the usual solutions to the Dirac equation, and $\epsilon_\mu$ is the massive polarization vector, with $m_V$ the mass of the spin 1 particle. Considering a Yukawa interaction of the form
\al{
\La_\text{Yukawa} = -\phi \left(g_L^{\phantom{\dagger}} \psi_R^\dagger \psi_L^{\phantom{\dagger}} +g_R^{\phantom{\dagger}} \psi_L^\dagger \psi_R^{\phantom{\dagger}}\right),
}
the corresponding on-shell amplitude with all-in convention is
\al{
{\cal A}_3\!\left[ \phi\psi_1^I \bar \psi_2^J\right] 
& = -\bar u^J(-p_2)\begin{pmatrix} g_L & 0\\ 0 & g_R \end{pmatrix} u^I(p_1) = g_L \braket{1^I2^J} + g_R \sbraket{1^I2^J}.
}
For vectors the relevant interactions are
\be
\La_V = \frac{m_Vc_0}{2} \phi V_\mu V^\mu + g\phi V_{\mu\nu}V^{\mu\nu} + \tilde{g}\phi V_{\mu\nu}\tilde{V}^{\mu\nu}.
\ee
The corresponding amplitude is
\al{\label{eq:phiVV_QFT}
\Amp_3\!\left[\phi V_1^{I_1,I_2}V_2^{J_1,J_2}\right] &=\left(2\frac{m_Vc_0}{2}\eta_{\mu\nu}+4g[p_1^\nu p_2^\mu -(p_1\cdot p_2)\eta_{\mu\nu}]+4\tilde g \epsilon_{\mu\nu\alpha\beta}p_1^\alpha p_2^\beta \right)\epsilon^\mu(p_1)\epsilon^\nu(p_2)\\
& = 2(i\tilde g - g)\braket{\bm{12}}^2-2(i\tilde g+g)\sbraket{\bm{12}}^2 -\frac{c_0}{m_V}\braket{\bm{12}}\sbraket{\bm{12}},
}
where we note explicitly the $1/m_V$ scaling of the last term. To manipulate the Levi--Civita we have used
\be
\epsilon_{\mu\nu\alpha\beta}=\frac{1}{4i}\Big(\Tr[\bar{\sigma}_\mu\sigma_\nu\bar{\sigma}_\alpha\sigma_\beta]-\Tr[\sigma_\mu\bar{\sigma}_\nu\sigma_\alpha\bar{\sigma}_\beta]\Big).
\ee

\subsection{SM amplitudes}

For completeness, we present here as well the SM 3-point amplitudes we use to derive our results in Section~\ref{sec:SM}:
\al{\label{eq:SM_amps}
\Amp_3\!\left[Q^{ibn}_1\bar{u}^{a}_{m2} H^j\right]=(Y_u^\dagger)^{ab}\delta^n_m\epsilon^{ij}\braket{12},&\quad \Amp_3\!\left[\bar{Q}^{b}_{in1}u^{am}_2\bar H_j\right]=(Y_u^{\phantom{\dagger}})^{ba}\delta^m_n\epsilon_{ij}\sbraket{12},\\
\Amp_3\!\left[Q^{ibn}_1\bar{d}^a_{m2}\bar H_j\right]=(Y_d^\dagger)^{ab}\delta^n_m\delta^i_j \braket{12},&\quad \Amp_3\!\left[\bar{Q}^{b}_{in1}d^{am}_2H^j\right]=(Y_d^{\phantom{\dagger}})^{ba}\delta^m_n\delta^j_i\sbraket{12},\\
\Amp_3\!\left[Q^{ibn}_1 \bar Q^{a}_{jm2} G^{A+}_3 \right]=-\sqrt{2}g_s T^A_{mn}\delta^i_j\delta^b_a \frac{\sbraket{23}^2}{\sbraket{12}},&\quad \Amp_3\!\left[Q^{ibn}_1 \bar Q^{a}_{jm2} G^{A-}_3 \right]=\sqrt{2}g_s T^A_{mn}\delta^i_j\delta^b_a \frac{\braket{13}^2}{\braket{12}},\\
\Amp_3\!\left[u^{bn}_1 \bar u^{a}_{m2} G^{A+}_3 \right]=\sqrt{2}g_s T^A_{mn} \delta^b_a \frac{\sbraket{13}^2}{\sbraket{12}},&\quad \Amp_3\!\left[u^{bn}_1 \bar u^{a}_{m2} G^{A-}_3 \right]=-\sqrt{2}g_s T^A_{mn}\delta^b_a \frac{\braket{23}^2}{\braket{12}},\\
\Amp_3\!\left[H_{1}^i\bar H_{j2} W_3^{A-}\right] = - \sqrt{2}g T^A_{ji} \frac{\braket{13}\braket{23}}{\braket{12}} ,&\quad \Amp_3\!\left[H_{1}^i\bar H_{j2} W_3^{A+}\right] = - \sqrt{2}g T^A_{ji} \frac{\sbraket{13}\sbraket{23}}{\sbraket{12}},\\
\Amp_3\!\left[H_{1}^i\bar H_{j2} B_3^{-}\right] = - \sqrt{2}g'y_H \delta_{j}^i \frac{\braket{13}\braket{23}}{\braket{12}} ,&\quad \Amp_3\!\left[H_{1}^i\bar H_{j2} B_3^{+}\right] = - \sqrt{2}g'y_H \delta^i_j \frac{\sbraket{13}\sbraket{23}}{\sbraket{12}},
}
where the expressions are analogous for other fermions and vector bosons, $g',g,g_s$ are the gauge couplings and $y_H$ the Higgs hypercharge. In the equations above $a,b$ denote flavor indices, $i,j$ indices from the fundamental of $SU(2)_L$ and $m,n$ for the fundamental of $SU(3)_c$. We use $T^A$ for the generators of both non-abelian groups. Here, the conventions are chosen such to correspond to covariant derivatives defined as $D_\mu=\del_\mu - ig V_\mu$, and the Yukawas as $\La_\text{Yukawa} = - Y_\psi \bar \psi_L H \psi_R+h.c.$.

\section{Yukawa - Mass connection}\label{app:Yukawa-mass}

In this Appendix we present an alternative way to understand the relation between fermion masses in the broken electroweak phase and the Yukawa matrices in the unbroken phase. This relation was previously studied in Ref.\,\cite{Balkin:2021dko} (and more recently in Ref.\,\cite{Bachu:2023fjn}) by analysing how the ``freezing'' of external Higgs particles can make massless spinors become massive. Here instead, we show explicitly that we can understand the Higgsing as a modification to the dispersion relation of the fermions, which is very similar to the analysis performed in Ref.\,\cite{Alves:2021rjc} for neutrino propagation in matter.

Consider a generic amplitude $\Amp$ that has a factorization channel in $\psi_R$ for $p^2\to 0$, with $\psi_R$ being one of the massless right-handed fermions of the SM. More precisely,
\be
\lim_{p^2\to 0}\Amp= \lim_{p^2\to0}\Amp_L^a\frac{\delta^{ab}}{p^2}\Amp_R^b = 
\adjustbox{valign=m}{
\begin{tikzpicture}[line width=0.75]
\draw[fill=white] (0,0) circle (15pt) node[midway] {$\Amp_L$};
\draw[f] (1-0.49,0) -- (3-0.49,0) node[midway, above] {$\psi_R$};
\draw[fill=white] (3,0) circle (15pt) node(3,0) {$\Amp_R$};
\end{tikzpicture}
} 
,
\ee
where $\Amp_{L,R}^a$ are the sub-amplitudes, $a,b$ are flavor indices and we leave implicit spinorial indices and the Levi--Civita tensor. Consider now the same amplitude with the addition of an Higgs - anti-Higgs pair in the $p_H,p_{\bar H}\to 0$ limit, in such a way that the Higgses become non-dynamical. Given that the Higgs interacts with the fermions through Yukawa interactions,
\be
\Amp_3\!\left[\psi_{L,1}^b\bar \psi_{R,2}^a \bar H\right] = (Y_\psi^\dagger)^{ab}\braket{12},\quad \Amp_3\!\left[\bar \psi_{L,1}^a\psi_{R,2}^b  H\right] = (Y_\psi^{\phantom{\dagger}})^{ab}\sbraket{12},
\ee
where $H\leftrightarrow \bar H$ in the case of $u$-quarks and ignoring SM group indices, we can make recursive use of \textit{polology} to show that the amplitude will feature extra factorization channels as $p^2,p_H,p_{\bar H}\to 0$. Schematically\footnote{Here we ignore the contributions from the Higgs being emitted by the other external legs, that contribute to them becoming massive and therefore are not relevant to our computations.},
\be
\lim_{p_H,p_{\bar H}\to 0}\lim_{p^2\to 0} \Amp\!\left[H,\bar H\right] = 
\adjustbox{valign=m}{
\begin{tikzpicture}[line width=0.75]
\draw[fill=white] (0,0) circle (15pt) node[midway] {$\Amp_L$};
\draw[f] (1-0.49,0) -- (2-0.49,0) node[midway, above] {$\psi_R$};
\filldraw[black] (2-0.49,0) circle (2pt);
\filldraw[black] (2-0.49,-0.75) circle (2pt);
\draw[dashed] (2-0.49,0) -- (2-0.49,-0.75) node[below] {$H$};
\draw[f] (2-0.49,0) -- (3-0.49,0) node[midway, above] {$\psi_L$};
\filldraw[black] (3-0.49,0) circle (2pt);
\filldraw[black] (3-0.49,-0.75) circle (2pt);
\draw[dashed] (3-0.49,0) -- (3-0.49,-0.75) node[below] {$\bar H$};
\draw[f] (3-0.49,0) -- (4-0.49,0) node[midway, above] {$\psi_R$};
\draw[fill=white] (4,0) circle (15pt) node(4,0) {$\Amp_R$};
\end{tikzpicture}
} .
\ee
Note that we need one Higgs and one anti-Higgs, because otherwise we would not obtain a factorization to the same amplitudes $\Amp_{L,R}^a$. Evaluating the amplitude above we obtain
\al{\label{eq:appbaux}
&\lim_{p_H,p_{\bar H}\to 0}\lim_{p^2\to 0} v^2\Amp\!\left[H,\bar H\right]\\
&= \lim_{p_H,p_{\bar H}\to 0}\lim_{p^2\to 0}\Amp_L^a \frac{1}{p^2} \frac{v (Y_\psi^{\phantom{\dagger}})^{ca}\sbraket{(-p-p_H)p}}{(p+p_H)^2}\frac{v (Y_\psi^\dagger )^{bc}\braket{(p+p_H)(-p-p_H-p_{\bar H})}}{(p+p_H+p_{\bar H})^2}\Amp_R^b,
}
where we insert one factor of $v$ for each Higgs becoming non-dynamical. We remark that the momenta $p+p_H$ and $p+p_H+p_{\bar H}$ in the expression above are in principle massive and should carry a massive little-group index. We have checked that, since we take the limit $p_{H,\bar H}\to 0$ in the end, we obtain the same results using massless spinors. The numerator of Eq.\,\eqref{eq:appbaux} can be rewritten as
\be
\sbraket{(-p-p_H)p}\braket{(p+p_H)(-p-p_H-p_{\bar H})} \simeq \sbraket{pp_H}\braket{p_Hp}=(p+p_H)^2,
\ee
where we have neglected higher-order terms in Higgs momenta. With the simplification above the limit becomes
\be\label{eq:HH_mass_yukawa}
\lim_{p_H,p_{\bar H}\to 0}\lim_{p^2\to 0} v^2\Amp\!\left[H,\bar H\right]=\lim_{p^2\to 0}\Amp_R^b \frac{1}{p^2} \frac{v^2(Y_\psi^\dagger Y_\psi^{\phantom{\dagger}})^{ba}}{p^2}\Amp_L^a.
\ee
It is clear that Eq.\,\eqref{eq:HH_mass_yukawa} contributes to the original amplitude without the Higgs - anti-Higgs pair, as the latter are removed from the external states. Hence, after we take all Higgs non-dynamical we arrive at
\be
\lim_{p_H,p_{\bar H}\to 0}\lim_{p^2\to 0}\left( \Amp+v^2\Amp\!\left[H,\bar H\right]+\cdots\right)=\lim_{p^2\to 0}\Amp_R^b \frac{1}{p^2}\left[\delta^{ba}+ \frac{v^2(Y_\psi^\dagger Y_\psi^{\phantom{\dagger}})^{ba}}{p^2}+\cdots\right]\Amp_L^a,
\ee
where the dots denote similar amplitudes computed with more Higgs insertions, which are given by analogous expressions. It is possible to resum all the contributions:
\be
\frac{1}{p^2}\left[\delta^{ba}+ \frac{v^2(Y_\psi^\dagger Y_\psi^{\phantom{\dagger}})^{ba}}{p^2}+\cdots\right] = \left[\frac{1}{p^2-v^2Y_\psi^\dagger Y_\psi^{\phantom{\dagger}}}\right]^{ba},
\ee
hence it is clear that the Higgs "background" modifies the dispersion relations of the fermions by inducing an effective mass. Note that the propagator above does not respect locality manifestly, due to the matrix structure of the Yukawas. Nevertheless, it is always possible to find basis in which the propagator becomes diagonal and we therefore restore manifest locality. The necessity to perform such rotation is nothing but a realization of the mismatch between the massless and massive flavor basis. The mass matrix is given by
\be\label{eq:yukawa2_mass2}
M_\psi^2 = v^2 U_R^\dagger Y_\psi^\dagger Y_\psi^{\phantom{\dagger}} U_R^{\phantom{\dagger}} = v^2\left(Y_\psi^\dagger Y_\psi^{\phantom{\dagger}}\right)_\text{diagonal},
\ee
where $U_R\in U(3)_{\psi_R}$, the right-handed flavor group of the massless phase of the theory. To be able to write directly the mass matrix in terms of the Yukawa, we can repeat the same steps above with a factorization channel on a left-handed fermion and obtain similarly $M_\psi^2 =U_L^\dagger Y_\psi^{\phantom{\dagger}} Y_\psi^\dagger U_L^{\phantom{\dagger}}$, where now $U_L\in U(3)_{\psi_L}$. In order for both expressions to match, the mass must be given by
\be\label{eq:yukawa_mass}
M_\psi^{\phantom{\dagger}} = v U_L^\dagger Y_\psi^{\phantom{\dagger}} U_R^{\phantom{\dagger}}.
\ee
It is interesting to notice that for the neutrino sector, in the absence of a Yukawa interaction involving right-handed neutrinos, we can still generate masses via the Weinberg amplitude
\be
\Amp_4\!\left[H^2L^2\right] = c_W \braket{12},
\ee
with $L$ the lepton doublet. Repeating the reasoning above for this interaction, we obtain
\be
M_\nu^2 = v^4(c_W^\dagger c_W^{\phantom{\dagger}})_\text{diagonal},
\ee
and we can thus see explicitly the different scaling with the scale $v$ when compared to Eq.\,\eqref{eq:yukawa_mass}.

\section{One-loop running of $\Amp[\phi \psi\bar{\psi}H]$}\label{app:running}

In this Appendix we describe in more detail the computation of the beta function used in Section~\ref{sec:SM} using the methods of generalized unitarity\,\cite{Zwiebel:2011bx,Wilhelm:2014qua, Caron-Huot:2016cwu,Forde:2007mi,Mastrolia:2009dr,Giele:2008ve,Bern:1994cg,Bern:2004cz,Britto:2004nc,Bern:1994zx,Abreu:2022mfk,Blumlein:2022zkr,Baratella:2020lzz,EliasMiro:2020tdv,Jiang:2020mhe,Henn:2014yza}. For 1-loop beta functions, the master formula reads\,\cite{Mastrolia:2009dr,Baratella:2020lzz}
\be\label{eq:1loop_master}
\left(\gamma_{\mathcal{O}_i}-\gamma_{\rm IR}^{(\mathcal{O}_i)}\right)\Amp\left[\mathcal{O}_i\right] = \frac{i}{8\pi^4}\sum_a\int \dd\Pi[\ell_1,\ell_2]\oint_{z=\infty}\frac{\dd z}{z} \Amp_L^{(a)}\!\left[\hat{\ell}_1,\hat{\ell}_2\right]\Amp_R^{(a)}\!\left[-\hat{\ell}_1,-\hat{\ell}_2\right],
\ee
where $\gamma_{\mathcal{O}_i}$ is the anomalous dimension of the amplitude $\Amp\left[\mathcal{O}_i\right]$. The formula above is obtained by selecting bubble diagrams from the Passarino--Veltman decomposition\,\cite{Passarino:1978jh}, which is the only topology that can contain UV divergences. This is achieved by first performing all 2-cuts (labeled by $a$), that result in the product of two tree amplitudes $\Amp^{(a)}_{L,R}$ integrated over the 2-body phase-space $\dd\Pi[\ell_1,\ell_2]$ of the two cut momenta $\ell_{1,2}$. Doing so does not select only bubbles, since 2-cuts get contributions from triangles and boxes as well. To remove them, we shift $\ell_{1,2}$ to the complex plane using a BCFW-like shift\,\cite{Britto:2004ap,Britto:2005fq}, $\hat \ell_{1,2}=\ell_{1,2}\pm z \ell_{2,1}$ and integrate over $\dd z/z$ around $z=0$. Then, we deform the contour around the origin as a sum over contours around poles and at $z=\infty$. The poles can only come from triangles and boxes, as they have uncut propagators. So dropping them and keeping only the residue at infinity guarantees that we are selecting only the bubbles. The numerical factor $i/8\pi^4$ arises from collecting the divergent piece of the bubble ($-1/8\pi^2$), normalizing the phase-space integral ($2/\pi$) and from Cauchy's theorem ($1/2\pi i$). Also, $\gamma_\text{IR}^{(\mathcal{O}_i)}$ denotes the IR contribution to the anomalous dimension that must be subtracted, and that depends only on the external particles of $\Amp\left[\mathcal{O}_i\right]$.

We are interested in computing the contribution of the amplitude $\Amp_3\!\left[\phi G G\right]$ to the running of $\Amp_4\!\left[\phi\psi \bar \psi H\right]$. For concreteness, let us choose $\Amp_4\!\left[\phi \bar Q d H \right]$ and $\Amp_4\!\left[\phi Q \bar d \bar H\right]$. For the first amplitude we have only two possible 2-cuts:
\be\label{eq:2cuts}
(\text{I})=
\adjustbox{valign=m}{
\begin{tikzpicture}
\filldraw[black] (0,0) circle (2pt);
\draw[f] (0,0) -- (-1,-1) node[below] {$\bar Q^a_{in}$};
\draw[f] (1,-1) node[below] {$Q^{a'i'n'}$} -- (0,0);
\draw[dashed] (0,0) -- (-1,1) node[above] {$\phi$};
\draw[g] (0,0) -- (1,1) node[above] {$G^A$};
\filldraw[black] (2+1.2,0) circle (2pt);
\draw[f] (2+1.2,0) -- (1+1.2,-1) node[below] {$\bar Q^{b'}_{j'm'}$};
\draw[f] (3+1.2,-1) node[below] {$d^{bm}$} -- (2+1.2,0);
\draw[g] (2+1.2,0) -- (1+1.2,1) node[above] {$G^{A'}$};
\draw[sb] (2+1.2,0) -- (3+1.2,1) node[above] {$H^j$};
\draw[dashed] (1.6,1.5) -- (1.6,-1.5);
\end{tikzpicture}},
\quad
(\text{II})=
\adjustbox{valign=m}{
\begin{tikzpicture}
\filldraw[black] (0,0) circle (2pt);
\draw[f] (-1,-1)  node[below] {$d^{bm}$} -- (0,0);
\draw[f] (0,0) -- (1,-1) node[below] {$\bar d^{b'}_{m'}$};
\draw[dashed] (0,0) -- (-1,1) node[above] {$\phi$};
\draw[g] (0,0) -- (1,1) node[above] {$G^A$};
\filldraw[black] (2+1,0) circle (2pt);
\draw[f] (1+1,-1) node[below] {$d^{a'}_{n'}$} -- (2+1,0);
\draw[f] (2+1,0) -- (3+1,-1) node[below] {$\bar Q^a_{in}$};
\draw[g] (2+1,0) -- (1+1,1) node[above] {$G^{A'}$};
\draw[sb] (2+1,0) -- (3+1,1) node[above] {$H^j$};
\draw[dashed] (1.5,1.5) -- (1.5,-1.5);
\end{tikzpicture}
}
\ee
where we ignore other cuts that do not involve $\Amp_3\!\left[\phi G G\right]$. In addition, $A,A'$ and $n,n',m,m'$ are indices of the adjoint and fundamental of $SU(3)_c$, respectively, $a,a',b,b'$ of flavor and $i,i',j,j'$ of the fundamental of $SU(2)_L$. To compute the 2-cuts above, we need the corresponding 4-point amplitudes. They can be computed using Eqs.\,\eqref{eq:ALP_2gluons_SM} and \eqref{eq:SM_amps}:
{\allowdisplaybreaks
\begin{align}\label{eq:4point_amplitudes}
\Amp_4\!\left[\phi Q_1^{a'i'n'}\bar Q^a_{in2}G^{A+}_3\right] = -\frac{\sqrt{2}g_sg_G^+}{f}T^A_{nn'}\delta^{i'}_i\delta^{aa'}\frac{\sbraket{23}^2}{\sbraket{12}},\nonumber\\
\Amp_4\!\left[\phi d_1^{a'n'}\bar d^a_{i2}G^{A+}_3\right] = \frac{\sqrt{2}g_sg_G^+}{f}T^A_{nn'}\delta^{aa'}\frac{\sbraket{13}^2}{\sbraket{12}},\nonumber\\
\Amp_4\!\left[\phi Q_1^{a'i'n'}\bar Q^a_{in2}G^{A-}_3\right] = \frac{\sqrt{2}g_sg_G^-}{f}T^A_{nn'}\delta^{i'}_i\delta^{aa'}\frac{\braket{13}^2}{\braket{12}},\nonumber\\
\Amp_4\!\left[\phi d_1^{a'n'}\bar d^a_{n2}G^{A-}_3\right] = -\frac{\sqrt{2}g_sg_G^-}{f}T^A_{nn'}\delta^{aa'}\frac{\braket{23}^2}{\braket{12}},\\
\Amp_4\!\left[\bar Q^a_{in1}d^{a'n'}_{2}G^{A-}_3H^{i'}\right] =-\sqrt{2}g_sT^A_{nn'}(Y_d^{\phantom{\dagger}})^{aa'}\delta^{i'}_i\frac{\sbraket{12}^2}{\sbraket{23}\sbraket{13}},\nonumber\\
\Amp_4\!\left[ Q^{ain}_{1}\bar d^{a'}_{n'2}G^{A+}_3 \bar H_{i'}\right] =-\sqrt{2}g_sT^A_{nn'}(Y_d^\dagger)^{a'a}\delta^{i}_{i'}\frac{\braket{12}^2}{\braket{23}\braket{13}},\nonumber
\end{align}
}
where $T^A$ denotes the generators of $SU(3)_c$ and $g_G^\pm$ the ALP-gluon couplings. The product of the two amplitudes in the left 2-cut in Eq.\,\eqref{eq:2cuts} reads
\al{\label{eq:1st_2cut}
(\text{I})&=(-1)\Amp_4\!\left[\phi Q_{\ell_1}^{a'i'n'}\bar Q^a_{in1}G^{A+}_{\ell_2}\right]\Amp_4\!\left[\bar Q^{b'}_{j'm'(-\ell_1)}d^{bm}_{2}G^{A'-}_{(-\ell_2)}H^{j}_4\right]\delta^{AA'}\delta^{b'a'}\delta_{i'}^{j'}\delta^{m'}_{n'}\\
& = -\frac{2\mathcal{C}_A(\bm{3})g_s^2g_G^+}{f}(Y_d)^{ab}\delta_i^j\delta_n^m \frac{\sbraket{1\ell_2}^2\sbraket{\ell_12}^2}{\sbraket{1\ell_1}\sbraket{\ell_1\ell_2}\sbraket{2\ell_2}},
}
with $\mathcal{C}_A(\bm{3})$ the Casimir of the adjoint of $SU(3)_c$. The extra minus sign above takes into account fermion ordering, \textit{i.e.} to arrange the amplitudes as we have defined in Eqs.\,\eqref{eq:SM_amps} and \eqref{eq:4point_amplitudes} we need to anti-commute some fermions, leading to an extra minus. Note that only one helicity configuration of the gluons contribute, as the SM amplitude $\Amp_4\!\left[\bar Q d GH\right]$ is non-zero for only one choice of gluon helicity. Then, we shift $\ell_{1,2}$ to the complex plane as
\be
\ket{\ell_1}\to \ket{\ell_1}+z\ket{\ell_2},\quad \sket{\ell_2}\to \sket{\ell_2}-z\sket{\ell_1},
\ee
while $\sket{\ell_1}$ and $\ket{\ell_2}$ remain unchanged. After performing the shift above and selecting the residue at $z=\infty$ we obtain
\al{\label{eq:2cut_after_residue}
(\text{I}) \to -2\pi i\frac{2\mathcal{C}_A(\bm{3})g_s^2g_G^+}{f}(Y_d)^{ab}\delta_i^j\delta_n^m \frac{2\sbraket{1\ell_2}\sbraket{2\ell_1}-\sbraket{1\ell_1}\sbraket{2\ell_2}}{\sbraket{\ell_1\ell_2}}.
}
Now comes the phase-space integration. To this end, we write $\ell_{1,2}$ as linear combinations of external momenta\,\cite{Zwiebel:2011bx,Caron-Huot:2016cwu,Balkin:2021dko,EliasMiro:2020tdv},
\be\label{eq:loop_momenta_matrix}
\begin{pmatrix} \sket{\ell_1} \\ \sket{\ell_2} \end{pmatrix} = \begin{pmatrix} \cos\theta & -e^{i\phi}\sin\theta \\ e^{-i\phi}\sin\theta & \cos\theta \end{pmatrix}\begin{pmatrix} \sket{4} \\ \sket{2} \end{pmatrix},
\ee
where $\phi$ is the azimutal and $\theta$ the half-polar angles. The angle spinors are related in similar way, but with an extra complex conjugation. Inserting Eq.\,\eqref{eq:loop_momenta_matrix} in \eqref{eq:2cut_after_residue} and integrating over phase-space
\be
\int \dd \Pi[\ell_1,\ell_2]~ (\text{I}) = -2\pi i\frac{2\mathcal{C}_A(\bm{3})g_s^2g_G^+}{f}(Y_d)^{ab}\delta_i^j\delta_n^m \left(-\frac{3\pi}{4}\sbraket{12}\right).
\ee
The 2-cut on the right in Eq.\,\eqref{eq:2cuts} is computed analogously. The product of the amplitudes is
\al{
(\text{II}) &= \Amp_4\!\left[\phi d^{bm}_2 \bar d^{b'}_{m'\ell_1}G^{A+}_{\ell_2}\right]\Amp_4\!\left[\bar Q^a_{in1}d^{a'}_{n'(-\ell_1)}G^{A'-}_{(-\ell_2)}H^j_4\right]\delta^{AA'}\delta^{b'a'}\delta_{m'}^{n'}\\
&= \frac{2\mathcal{C}_A(\bm{3})g_s^2g_G^+}{f}(Y_d)^{ab}\delta_i^j\delta_n^m \frac{\sbraket{2\ell_2}^2\sbraket{\ell_11}^2}{\sbraket{2\ell_1}\sbraket{\ell_1\ell_2}\sbraket{1\ell_2}},
}
where now we do not have any extra minus from fermion ordering. Note that the spinor structure from expression above is identical to the one in Eq.\,\eqref{eq:1st_2cut} by changing $1\leftrightarrow 2$. Thus,
\be
\int \dd \Pi[\ell_1,\ell_2]~ (\text{II}) = 2\pi i\frac{2\mathcal{C}_A(\bm{3})g_s^2g_G^+}{f}(Y_d)^{ab}\delta_i^j\delta_n^m \left(\frac{3\pi}{4}\sbraket{12}\right).
\ee
The result of both 2-cuts must be then inserted in Eq.\,\eqref{eq:1loop_master} and compared to\linebreak $\Amp_4\!\left[\phi \bar Q d H\right]$. It is important to notice that the 2-cuts produced the same kinematical structure as the original amplitude. Besides, there is no contribution from IR divergences in this case\,\cite{Machado:2022ozb,AccettulliHuber:2021uoa}. We arrive at
\be\label{eq:anomalous_dim}
\frac{\dd C_d}{\dd\log\mu} = -\frac{g_s^2g_G^+}{\pi^2}Y_d^{\phantom{\dagger}},\quad \frac{\dd \bar C_d}{\dd\log\mu} = -\frac{g_s^2g_G^-}{\pi^2}Y_d^\dagger,
\ee
where we included the result also for the conjugate amplitude, that follows in a very similar way. The results above, properly translated to the usual language by Eq.\,\eqref{eq:phiVV_QFT}, agree with previous computations in the literature\,\cite{Chala:2020wvs,Bauer:2020jbp,DasBakshi:2023lca}. At leading order, the couplings $g_G^\pm$ are related by complex conjugation according to Eq.\,\eqref{eq:CPT_amp}, which implies that
\be
\frac{\dd \bar C_d}{\dd\log\mu}\simeq \frac{\dd C_d^\dagger}{\dd\log\mu},
\ee
up to two-loop effects. As a consequence, we can use the results of Section~\ref{sec:SM} and, to be consistent with the 1-flavor limit, must satisfy $g_G^-=-g_G^+$. Therefore,
\be
g_G^-=-g_G^+=iC_{\phi GG},\quad C_{\phi GG}\in \mathbb{R},
\ee
that corresponds solely to the $\phi G\tilde G$ coupling, as we wanted to demonstrate.

\bibliographystyle{JHEP2}
\bibliography{draft_PNGBs}

\end{document}